\documentclass[twocolumn]{openjournal}

\usepackage{orcidlink}
\usepackage{lineno}
\usepackage{comment}
\usepackage{color}
\usepackage{hyperref}
\hypersetup{
	colorlinks=true,
	urlcolor=blue,    
	linkcolor=black,  
	citecolor=blue,   
}
\newcommand{\orcidauthor}[3]{\author{\href{http://orcid.org/#1}{#2$^{#3}$}}}

\usepackage{amsmath}

\shorttitle{Rotating Supergiants from Binary Merger Products}
\shortauthors{Tsuna, Fuller, Lu}

\begin{document}

\title{Fates of Rotating Supergiants from Stellar Mergers and the Landscape of Transients upon Core-collapse\vspace{-1.6cm}}

\orcidauthor{0000-0002-6347-3089}{Daichi Tsuna}{1,2,*}
\orcidauthor{0000-0002-4544-0750}{Jim Fuller}{1}
 \orcidauthor{0000-0002-1568-7461}{Wenbin Lu}{3}
 \affiliation{$^{1}$TAPIR, Mailcode 350-17, California Institute of Technology, Pasadena, CA 91125, USA}
 \affiliation{$^{2}$Research Center for the Early Universe, School of Science, The University of Tokyo, 7-3-1 Hongo, Bunkyo, Tokyo 113-0033, Japan}
\affiliation{$^{3}$Department of Astronomy and Theoretical Astrophysics Center, University of California, Berkeley, CA 94720-3411, USA}

\thanks{$^*$E-mail: \href{mailto:tsuna@caltech.edu}{tsuna@caltech.edu}}


 \begin{abstract}
We present a grid of rotating supergiant models from post-main sequence binary merger products, constructed by the MESA stellar evolution code. We focus on the evolution of these stars until core-collapse, in addition to their rotation, which could influence their mass loss and explosion phenomenology. 
We find that (i) as in previous studies, larger mass gain by merger favors the production of blue supergiants (BSGs) over red supergiants, and (ii) merger products that end as BSGs at core collapse have rotating outer envelopes, with lower-mass BSGs having faster envelope rotation due to less wind mass loss. We model the expected transients from these BSGs upon core-collapse, considering cases where the neutrino-driven explosion is successful and unsuccessful. The successful explosions result in supernovae (SNe) with long-rising light curves of morphology similar to SN 1987A. Failed explosions of these BSGs result in envelope fallback of $\sim (0.1$-- several) $~M_\odot$  over $10^3$--$10^5$ seconds that power strong ($10^{51}$--$10^{53}$ erg) accretion-driven outflows in winds and possibly jets, with relativistic jets (if formed) generally capable of breaking out of the BSG envelope. Our modeling points to these merger-origin BSGs as viable progenitors for SN 1987A-like SNe, ultra-long gamma-ray bursts, and some of the fast luminous transients found in high-cadence optical surveys.
 \end{abstract}


\section{Introduction}
Massive stars are observed in the local universe in large numbers, occupying diverse regions of the Hertzsprung-Russell (HR) diagram. The diversity is shaped by various processes in the star, such as nuclear burning, convection, radiation transport and mass loss. These stars also often exist in binaries or higher-order multiples \citep{Duchene13,Sana14}, and binary interactions can drastically affect the evolutionary paths throughtout their lives \citep{Sana12}. 

In the local universe, there has been a long-standing puzzle on the abundant population of blue supergiants (BSGs; e.g., \citealt{Humphreys84,Fitzpatrick90,Castro14,Patrick25}). They reside in a region of the HR diagram where single star evolution theory predicts a gap (Hertzsprung gap), due to the rapid evolution of these stars to red supergiants (RSGs) once they leave the main sequence. Among the channels proposed for long-lived BSGs, a promising one is mergers/accretion in binaries \citep{Braun95,Glebbeek13,Vanbeveren13,Justham14,Farrell19,Menon24,Schneider24}, which generally leads to stars with undermassive helium cores and potentially higher surface helium/nitrogen abundance due to dredge-up of the burned layer to the surface. 

While there have been successful attempts in reproducing the BSG population via binary merger products, the final fates of these stars are less explored. The best-studied case is SN 1987A, an explosion of a BSG in the Large Magellanic Cloud \citep[LMC;][]{West87}. The equatorial ring seen in the supernova (SN) remnant phase supports the BSG progenitor being a product of a binary merger that occurred $\sim 10^4$ yrs before core-collapse \citep[e.g.,][]{Hillebrandt89,Podsiadlowski90,Podsiadlowski92,Morris09,Menon17,Urushibata18}. On the other hand, failed SNe of rotating BSGs result in fallback accretion onto the compact remnant, and have also been suggested as progenitors of various high-energy transients \citep{Woosley12,Dexter13,Kashiyama13,Nakauchi13,Kashiyama15,Perna18,Margutti19,Tsuna21}. While stellar mergers provide a natural pathway for forming rotating supergiants, detailed modelling of such transients from these merger products has not been systematically explored.

In this work we model rotating evolved supergiants from post-main sequence stellar mergers, aiming to obtain a comprehensive understanding of their final fates. We couple a 1D framework of constructing and evolving merger products with prescriptions for angular momentum (AM) transport inside the star. Using these stellar models, we consider diverse transients that can be expected from the deaths of these stars. We find that BSGs are produced by gain of a significant fraction of its mass due to accretion. Furthermore, the merger products tend to maintain rapid envelope rotation until core-collapse, which can produce not only 1987A-like SNe but also accretion-driven high-energy transients in case of a failed SN.

In Section \ref{sec:mesa_model} we first describe our modeling of the stellar merger products and their evolution towards core-collapse, and in Section \ref{sec:mesa_results} discuss the diverse evolution of the structure and rotation of these stars until core-collapse. We then discuss in Section \ref{sec:transients} the expected transients upon core-collapse using our stellar models, showing that these post-main sequence merger products are viable progenitors for various classes of transients, from 1987A-like SNe to engine-powered high-energy transients such as ultra-long gamma-ray bursts (GRBs) and fast luminous transients like AT2018cow. We conclude in Section \ref{sec:conclusion}, with directions for future studies.

\section{Stellar Evolution Models}
\label{sec:mesa_model}
We evolve massive stars with the open-source stellar evolution code MESA r23.05.1 \citep{Paxton11,Paxton13,Paxton15,Paxton18,Paxton19,Jermyn23}\footnote{A Zenodo link for the MESA inlists and the stellar models at core-collapse will be released upon acceptance of this manuscript.}. Throughout this work we fix the stars to a metallicity of $Z=0.007$, comparable to the environment of the LMC. This is also close to the mean/median metallicities inferred from observations of the explosion sites of local core-collapse SNe \citep[12+$\log_{10}({\rm O/H})\approx$ 8.35--8.5;][]{Galbany16,Pessi23}, and could be regarded as representative of the local core-collapse SN population.

We consider rigidly rotating stars at zero age main sequence (ZAMS), with surface rotation velocity of 20\% of the critical velocity. The resulting surface rotation speeds of $\sim 100$ km s$^{-1}$ are similar to those observed for OB stars in LMC \citep{Hunter08,Ramachandran18,Ramachandran19}. We evolve the rotation profile of the star based on recent prescriptions of AM transport \citep{Fuller19,Ma19}, with parameters adopted in \cite{Fuller22}. Mass loss is included under the Dutch wind prescription \citep{deJager88,Nugis00,Vink01}, with an efficiency (\verb|Dutch_scaling_factor|) of 0.5. The mass loss rate also depends on the surface rotation velocity, as described in Section 6.4 of \cite{Paxton13}.

For convection we adopt a mixing length parameter of $\alpha_{\rm MLT}=1.5$, and a Ledoux criterion with semi-convection efficiency parameter $\alpha_{\rm sc}=10$ motivated from the properties of post-main sequence massive stars in the Small Magellanic Cloud \citep{Schootemeijer19}\footnote{This high $\alpha_{\rm sc}$, close to adopting a Schwarzschild criterion, is inferred from the observed fraction of BSGs in the SMC \citep{Schootemeijer19}. 
A high $\alpha_{\rm sc}$ as in the Schwarzschild criterion is also argued from independent stellar modeling, as a consequence of entrainment of the semi-convective zone by the convective zone \citep[e.g.,][]{Anders22} and/or double-diffusive instabilities occurring in semi-convective regions \citep[e.g.,][]{Moore16}].}. Convective core overshooting is taken into account as an exponential scheme in MESA with parameters $(f,f_0)=(0.02, 0.005)$. We turn off overshooting after core helium depletion, by setting the parameter \texttt{min\_overshoot\_q} in MESA to be a value above unity.

We mimic the merger process in MESA by rapid mass accretion to the primary, as done in previous works \citep[e.g.,][]{Justham14,Rui21,Schneider24,Menon24}. We consider the most likely case of (early) Case B merger products, where merger occurs during when the star rapidly expands across the Hertzsprung gap \citep[see e.g. Figure 1 of][]{Justham14}. When the star expands to a radius $R_*=50\ R_\odot$, we add mass with a rate of $\dot{M}_{\rm add}\approx 10^{-2}\ M_\odot\ {\rm yr}^{-1}$ until it accretes the mass of the companion (using \texttt{max\_star\_mass\_for\_gain}=$M_2$). This choice of $R_*$ is motivated by the primary to retain most of its envelope upon merger, before it expands too much and becomes loosely bound such that it may be ejected by common envelope evolution \citep[e.g.,][]{Klencki21}. In Appendix \ref{app:model_variations} we verify by running additional MESA models that the final outcome is not strongly sensitive to the detailed choice of the radius at the onset of mass accretion.

To make the accretion process numerically stable in MESA, we initially add mass at a lower rate $\dot{M}_{\rm init}=10^{-5}\ M_\odot\ {\rm yr}^{-1}$, and then evolve $\dot{M}$ to asymptote to the desired $\dot{M}_{\rm add}$ when the star roughly accretes a mass of $\Delta M_{\rm crit}$. We prescribe $\dot{M}$ as a function of cumulative accreted mass $\Delta M$ by
\begin{equation}
    \dot{M} = {\rm max}\left[\dot{M}_{\rm init},  \frac{\dot{M}_{\rm add}}{1+(\Delta M_{\rm crit}/\Delta M)}\right]
\end{equation}
where we take $\Delta M_{\rm crit}=10^{-2}~M_\odot$. In MESA, mass being added is set to be the same specific entropy as that at the surface \citep{Paxton15} due to the short thermal timescale expected at the photosphere. We also assume that the composition of the added mass is same as that at the photosphere of the primary, which is almost identical to the composition at ZAMS (see \citealt{Justham14,Schneider24} and Section \ref{sec:conclusion} for discussions on this assumption). 

For completeness we consider mergers with mass ratios $q=M_2/M_1$ up to $0.8$, which may occur when the secondary expands and also fills its own Roche Lobe upon mass transfer from the primary \citep{Pols94,Justham14}. However, we note that a relatively high $q$ may result in stable mass transfer that avoid merger depending on the donor/accretor properties and uncertain stability boundary \citep[e.g.,][]{Temmink23,Henecco24,Ercolino24}. We assume that the entire secondary's mass $M_2$ is accreted, and no mass is lost from the system upon merger. If one assumes a fraction $f_{\rm acc}<1$ of the secondary mass ends up accreted onto the primary, our models are essentially equivalent to merging a secondary with mass $M_2/f_{\rm acc}$ (a similar argument also applies if a fraction of the primary’s mass is lost during the merger process). We prescribe the specific AM of the accreted material $j_{\rm acc}$ by the orbital AM being fully inherited to the merger product,
\begin{eqnarray}
    j_{\rm acc}&=&M_1\sqrt{\frac{GR_*}{M_1+M_2}}\nonumber \\
    &\sim& 7\times 10^{19}\ {\rm cm^2\ s^{-1}}\left(\frac{M_1}{15~M_\odot}\right) \left(\frac{M_1+M_2}{20~M_\odot}\right)^{\!\!-1/2}.
\label{eq:j_acc}
\end{eqnarray}
where $G$ is the gravitational constant and here $R_*=50~R_\odot$.

After accretion we run the models until core carbon depletion, defined as when the central carbon mass fraction drops to $10^{-3}$. Afterwards we expect the envelope structure in the MESA models to be mostly unchanged, as the time until core-collapse ($\lesssim 10$ yrs) is much less than the Kelvin-Helmholtz (KH) timescale of supergiants (order $10^2$--$10^3$ yrs). We adopt the \texttt{approx21\_cr60\_plus\_co56} nuclear reaction network, appropriate for modeling the star up to this evolutionary stage.

\begin{figure*}
    \centering
        \centering
        \begin{tabular}{cc}
          \includegraphics[width=0.5\linewidth]{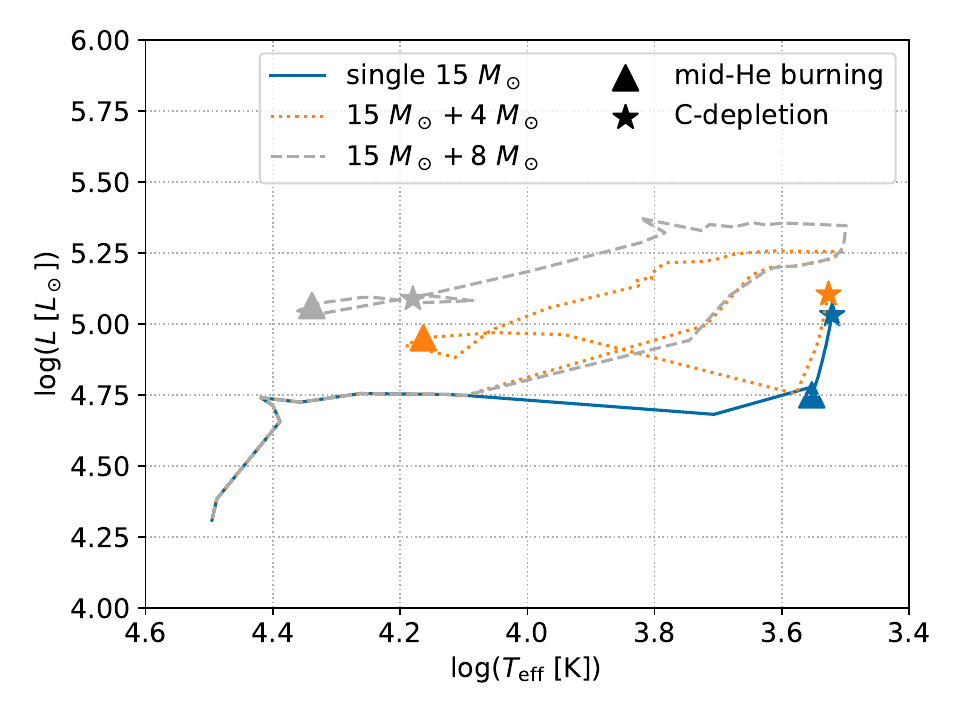}   &
          \includegraphics[width=0.5\linewidth]{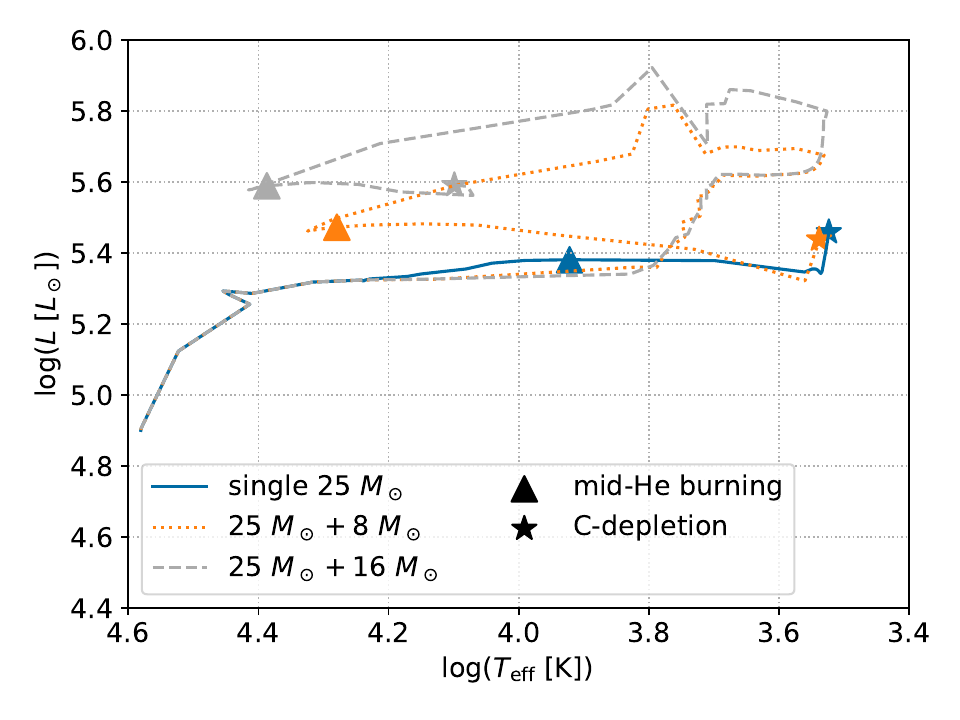}
        \end{tabular}
    \caption{Evolution in the HR diagram for representative stellar models, with initial mass of $15~M_\odot$ in the left panel and $25~M_\odot$ in the right panel. Triangles denote the locations of the models at the middle of core helium burning (defined as when $Y_{\rm He}=0.5$ in the center), and star symbols denote their locations at core carbon depletion where we expect them to remain until core-collapse. The tracks diverge where the star expands to $50~R_\odot$ post main sequence, when mass is added for the merger models.}
    \label{fig:HR_diagram}
\end{figure*}

We consider four primaries with initial masses of $M_1=[10,15, 20, 25]~M_\odot$, and accreted masses $M_2$ in steps of $2~M_\odot$ from $0~M_\odot$ (i.e. single star model) up to $0.8M_1$. This results in a total of 32 models, whose properties are summarized in Table \ref{tab:summary}. 

\section{Results}
\label{sec:mesa_results}

We first show evolutions of representative stellar models starting from a $15~M_\odot$ and $25~M_\odot$ primary to illustrate the underlying physics of the evolution of the surface properties and the AM profile. We then present the results for our grid of models to discuss the dependence of the merger product's fate on both $M_1$ and $M_2$.

\subsection{Evolution of Representative Models}
\label{sec:representative_models}

\subsubsection{Evolution on the HR Diagram}
\label{sec:HRdiagram_rep}

Figure \ref{fig:HR_diagram} shows the evolutionary tracks of the representative stellar models in the HR diagram. The solid lines in the two panels show the single star models, the dashed and dotted lines show merger models with different $M_2$, that ends its life either as a RSG or a BSG.

The single stars expand once they leave the main sequence, with slight differences in the detailed radial evolution for the $15~M_\odot$ and $25~M_\odot$ models. The former immediately expands and becomes a RSG as it crosses the Hertzsprung gap, while the latter expands more gradually and becomes a RSG near the end of core helium burning. Such variations in radial evolution with mass have been seen in previous post-main sequence stellar models for low metallicities \citep[e.g., Figures 1, C1 of][also \citealt{Georgy13,Schootemeijer19}]{Klencki20}. Despite these subtle differences with mass, all of our single star models eventually become RSGs well before core-collapse, with effective temperatures of $\approx 3300$ K and radii of $\approx 600$--$1600~R_\odot$.

The merger models have a more complicated evolution in the HR diagram, whose final fate depends on the mass $M_2$ being added to the star. The rapid accretion onto the primary, with accretion timescales much shorter than its KH timescale of the primary, initially inflates the envelope as the star cannot thermally adjust its structure \citep[e.g.,][]{Kippenhahn77,Neo77,Schurmann24,Lau24}. The detailed surface properties at this phase would be sensitive to the prescription for the mass accretion rate. 

We expect the envelope structure to be more robust from a few KH times after the end of mass accretion, when the merger product thermally relaxes. The merger models move to the blue side of the HR diagram over a timescale of $10^3$--$10^4$ years, and initiate core helium burning as a BSG. Models with low $M_2$, such as $15M_\odot+4M_\odot$ and $25M_\odot+8M_\odot$ in Figure \ref{fig:HR_diagram}, significantly expand soon after core helium depletion, spending the rest of their lives as a RSG. On the other hand, the models with more massive $M_2$, such as the $15M_\odot+8M_\odot$ and $25M_\odot+16M_\odot$ models, expand less after core helium depletion and remain as a BSG until core-collapse.

Such a behavior of stars turning blue with increasing mass accretion has been explained in the literature in multiple ways. The expansion of massive stars into RSGs have been attributed to the large pressure gradient established around the H-burning shell, when the core contracts soon after main sequence. A large instantaneous mass accretion increases the envelope mass (and gravity) while keeping the He core mass fixed, which makes the star harder to expand while maintaining hydrostatic equilibrium. Another picture was suggested in \cite{Schneider24}, where higher mass accretion increases the temperature at the base of the hydrogen envelope, turning the hydrogen shell-burning layer convective. We also observe the hydrogen shell-burning layer being convective for our BSG models of large $M_2$, with reduced pressure gradient around the shell. For both pictures, the resulting structure of such merger products is analogous to a less-evolved massive star in its main-sequence phase, explaining the compact, radiative envelope required for BSGs.

\subsubsection{Rotation Profile}
\label{sec:rotatin_rep}
\begin{figure*}
    \centering
    \includegraphics[width=\linewidth]{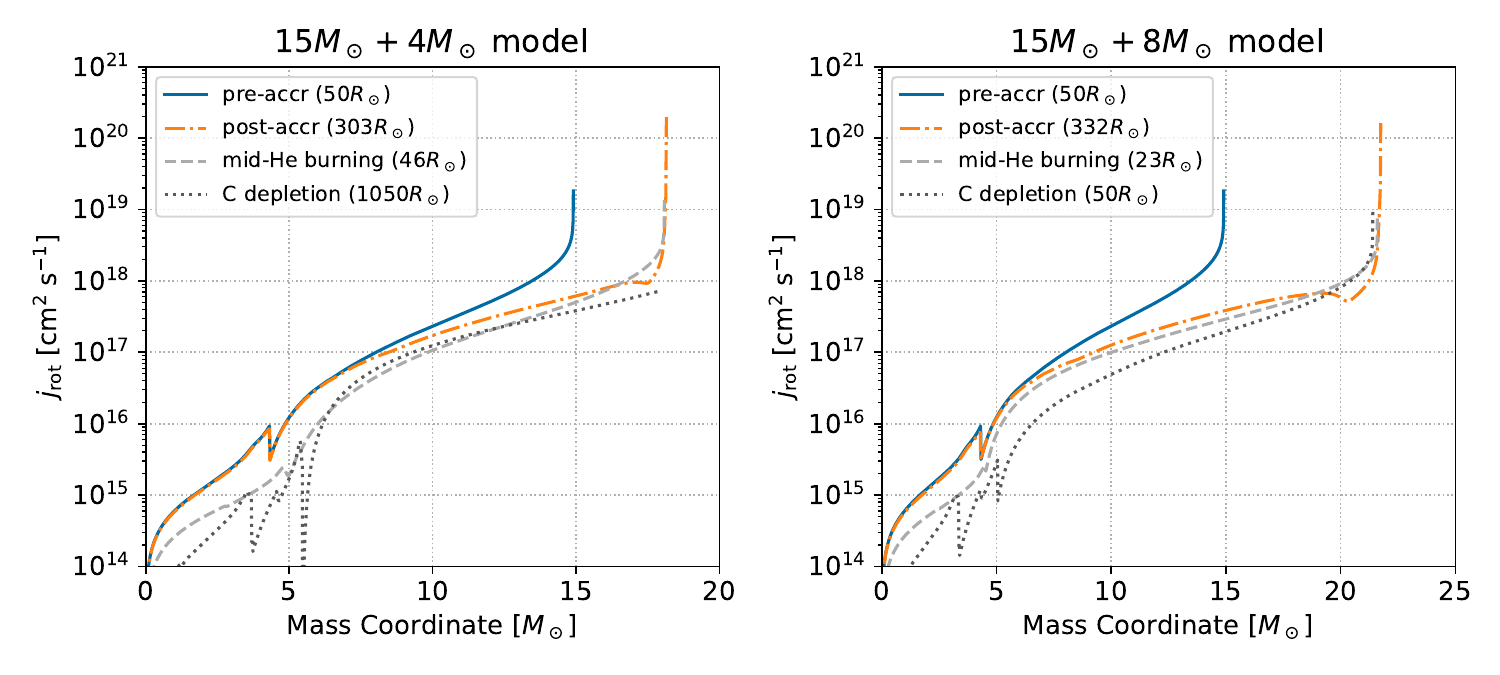} 
    \caption{Evolution of the specific AM for two merger models in Figure \ref{fig:HR_diagram} with $M_1=15\ M_\odot$. The lines denote the post main-sequence evolutionary phases of the stellar models, with their photospheric radii in parentheses. Solid lines are profiles just before mass accretion of the primary, and dash-dotted lines are those when the stars are contracting towards BSGs just after mass accretion. Dashed and dotted lines indicate the subsequent profiles at the middle of core helium burning and core carbon depletion.}
    \label{fig:j_rot}
\end{figure*}

Figure \ref{fig:j_rot} shows the specific AM profile for two representative merger product models, of a $15~M_\odot$ star accreting $4~M_\odot$ and $8~M_\odot$. The star before mass accretion has a profile sharply rising at the outermost mass coordinate. The sharp rise is expected for a radiative envelope with density sharply dropping with radius near the surface (i.e. $r$ rapidly increasing with increased $M_r$), and obeying rigid rotation (i.e. uniform $\Omega\approx j_{\rm rot}/(2/3)r^2$ -- note the weak dependence of the factor 2/3 on $\Omega$; \citealt{Paxton19}). AM transport arising from magnetic field amplification by Taylor instability \citep{Spruit02, Fuller19} is efficient in radiative zones with largest shear in composition interfaces. This realizes a nearly rigid rotation in the compositionally uniform H envelope, and discontinuities in the AM profile near the edge of the He core at a mass coordinate of around $4~M_\odot$.

After mass accretion the merger product begins to contract, and sheds mass due to the large AM gained from the merger. The critical specific AM at the surface where centrifugal force balances with gravity is
\begin{eqnarray}
    j_{\rm crit} &\approx& \frac{2}{3}\sqrt{\left(1-\Gamma_{\rm Edd}\right)GM_*R_*} \nonumber \\
    &\sim& 4\times 10^{20}\ {\rm cm^2\ s^{-1}}  \nonumber\\
    &&\times (1-\Gamma_{\rm Edd})^{1/2} \left(\frac{M_{\rm *}}{20M_\odot}\right)^{\!\!1/2} \left(\frac{R_*}{1500R_\odot}\right)^{\!\!1/2},
\end{eqnarray}
where $\Gamma_{\rm Edd}$ is the surface Eddington ratio which is well below unity ($0.1$--$0.5$) for our stellar models. As the star contracts to a BSG $j_{\rm crit}$ decreases to $\lesssim 10^{20}\ {\rm cm\ s^{-1}}$, resulting in the outer layers with $j>j_{\rm crit}$ lost as a (rotationally-enhanced) stellar wind. This results in a nearly instantaneous mass loss of typically $1$--$2~M_\odot$ in our models, and further mass loss by stellar winds brings the merger product to a sub-critical rotation. This kind of mass shedding accompanying envelope contraction has been suggested to be responsible for the equatorial circumstellar ring seen in SN 1987A \citep[][]{Morris07}, although for our stellar models the entire process occurs (by construction) much earlier than suggested for the progenitor of SN 1987A, at $\sim 10^6$ years before core-collapse. A qualitatively similar phenomena is also seen in simulations of main-sequence merger products \citep{Schneider19}.

For the $15~M_\odot + 4~M_\odot$ model, the subsequent expansion to a RSG after core helium depletion changes the AM profile to a flatter one, reflecting the flattening in the density structure of the envelope as it becomes convective. The AM profile of the $15~M_\odot+8~M_\odot$ model is nearly unchanged till core-collapse as the envelope remains radiative. 

While our prescriptions for AM evolution and transport do not capture the complicated three-dimensional processes that operate during the merger, our stellar models are found to be consistent with measured rotation rates of BSGs. The projected surface rotation velocities of observed BSGs in the LMC have a range of $v_{\rm surf}\sin i = (10$--$70)$ km s$^{-1}$ \citep{Menon24}. Taking an angle-average as a representative value,
    $\langle\sin i\rangle= \frac{1}{4\pi}\int_0^{2\pi}d\phi\int_0^{\pi} \sin i (\sin i)di = \frac{\pi}{4},$
the resulting observationally estimated surface AM is
\begin{eqnarray}
    j_{\rm surf} \sim (0.3{\rm -} 2)\times 10^{19}\ {\rm cm^2\ s^{-1}}\left(\frac{\langle\sin i\rangle}{\pi/4}\right)^{\!\!-1}\left(\frac{R_*}{50~R_\odot}\right),
\end{eqnarray}
consistent with our stellar models during core helium burning at an order of magnitude level.

\subsection{Model Grid}
\label{sec:model_grid}

\begin{figure*}
    \centering
    \includegraphics[width=\linewidth]{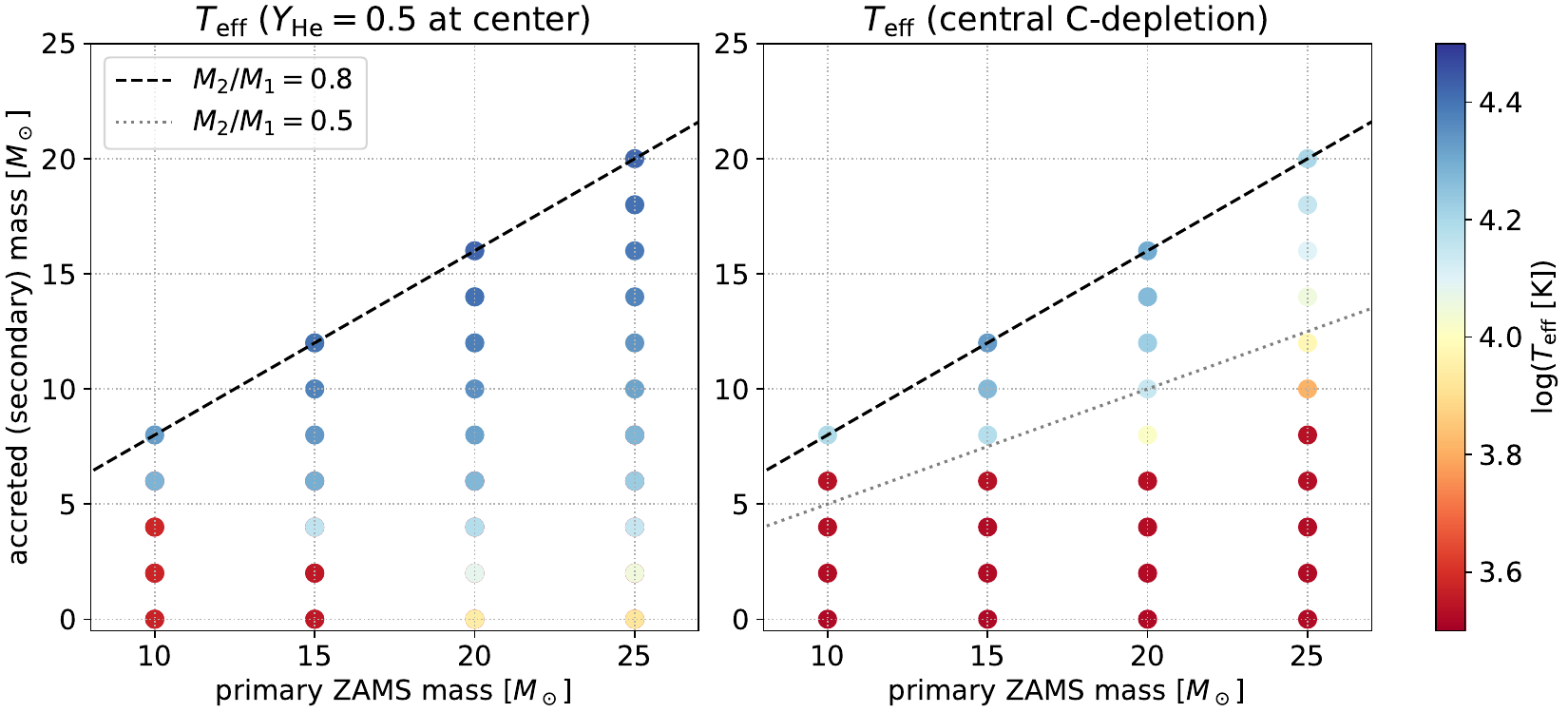}
    \caption{Effective temperatures of the simulated 32 progenitors, in the middle of core helium burning and at core carbon depletion. Colors denote the logarithm of the effective temperature. The dashed lines of $M_2/M_1=0.8$ denote our adopted upper limit where mergers are expected to occur \citep[e.g.,][]{Justham14}, and the dotted line of $M_2/M_1=0.5$ in the right panel shows the rough line separating the RSG/BSG populations at core-collapse.}
    \label{fig:Teff}
\end{figure*}

In Figure \ref{fig:Teff} we plot the effective temperature $T_{\rm eff}$ of the full stellar models at two stages of evolution, in the middle of core helium burning (with central helium mass fraction of $0.5$), and in the end of our simulations at core carbon depletion. The former is more relevant for the supergiants we observe locally (e.g. in the LMC), while the latter is relevant for the transients upon core-collapse.

We find the trend on $M_2$ for BSGs described in Section \ref{sec:HRdiagram_rep} to be generally true across the adopted range of $M_1$. When looked at the core He burning phase, the threshold of $M_2$ for forming a BSG is lower for larger primary mass. This reflects the delayed expansion for more massive stars at LMC metallicities described in Section \ref{sec:HRdiagram_rep}, with massive primaries only requiring a small amount of (or even zero) accretion to keep them blue during the helium burning phase. However, this trend is then reversed at core carbon depletion, where a larger $M_2$ is needed for forming BSGs for larger primary mass. With the exception of the $M_1=10~M_\odot$ models, the threshold for $M_2$ to form long-lived BSGs is roughly found as $M_2/M_1\gtrsim 0.5$, shown as a dotted line in the right panel of Figure \ref{fig:Teff}.

We compare our models with recent parameter studies that follow similar approaches to evolve merger-produced BSGs \citep{Schneider24,Menon24}. Our model qualitatively agrees with these works that larger accretion ($M_2$) more likely leads to BSGs, but our result lies somewhere in the middle of \cite{Schneider24} and \cite{Menon24} in terms of how much $M_2$ is required to create long-lived BSGs. In the range of $M_1$ we consider, we produce more BSGs than \cite{Schneider24} and less BSGs than \cite{Menon24}. A potential reason is that we consider a lower metallicity and higher $\alpha_{\rm sc}$ than \cite{Schneider24}, that both favor production of BSGs. On the other hand we neglect the dredge-up of He core and the hydrogen-burning shell during merger (as parameterized in e.g., \citealt{Menon24}), which is important for reproducing the surface abundances of helium and CNO elements in observed BSGs. While the detailed level of dredge-up driven by merger is highly uncertain for these systems, including this effect will make the merger product bluer due to the increased mean molecular weight in the envelope.

\begin{figure*}
    \centering
    \includegraphics[width=\linewidth]{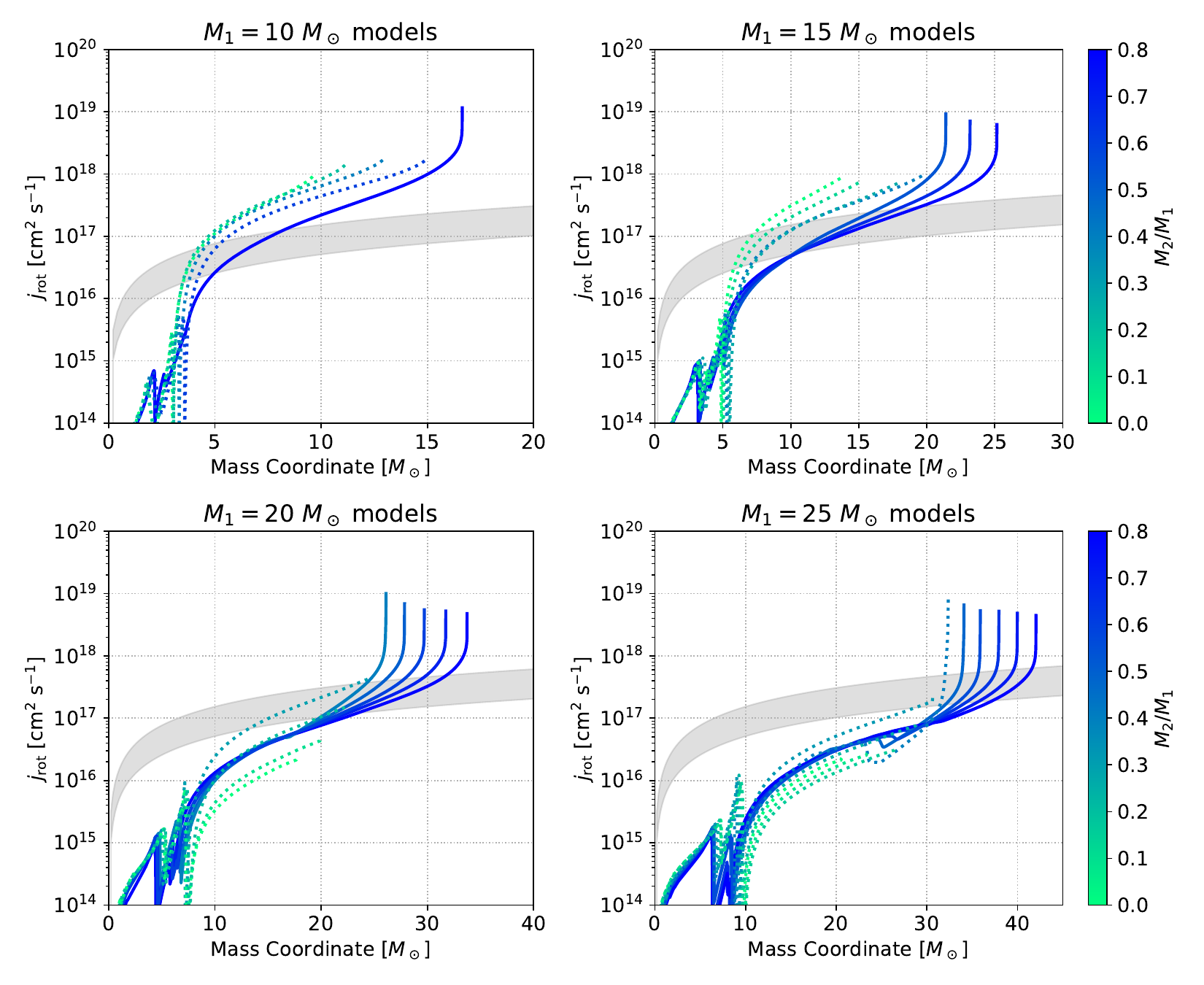}
    \caption{Final rotation profiles of the simulated 32 progenitors, with colors denoting $M_2/M_1$. Solid lines are those that end their lives as BSGs defined as $\log_{10}(T_{\rm eff, fin})>3.9$ (approximately A-type or hotter), and dashed lines are those that end as cooler supergiants. Gray regions show the specific AM required for material infalling onto the newborn BH to circularize outside the BH's ISCO (equation \ref{eq:j_ISCO}), which depends on the BH spin (being larger for lower BH spin).}
    \label{fig:jrot_all}
\end{figure*}

Figure \ref{fig:jrot_all} shows the AM profiles of the full stellar models at core carbon depletion, with solid and dotted lines indicating BSG of $T_{\rm eff}>10^{3.9}$~K and red/yellow supergiants of $<10^{3.9}$~K respectively. The cooler supergiants have a flatter rotation profile across the envelope, while the BSGs have a sharp rise near the surface as expected for a radiative envelope with nearly rigid rotation (Section \ref{sec:rotatin_rep}). 

\begin{figure}
    \centering   \includegraphics[width=\linewidth]{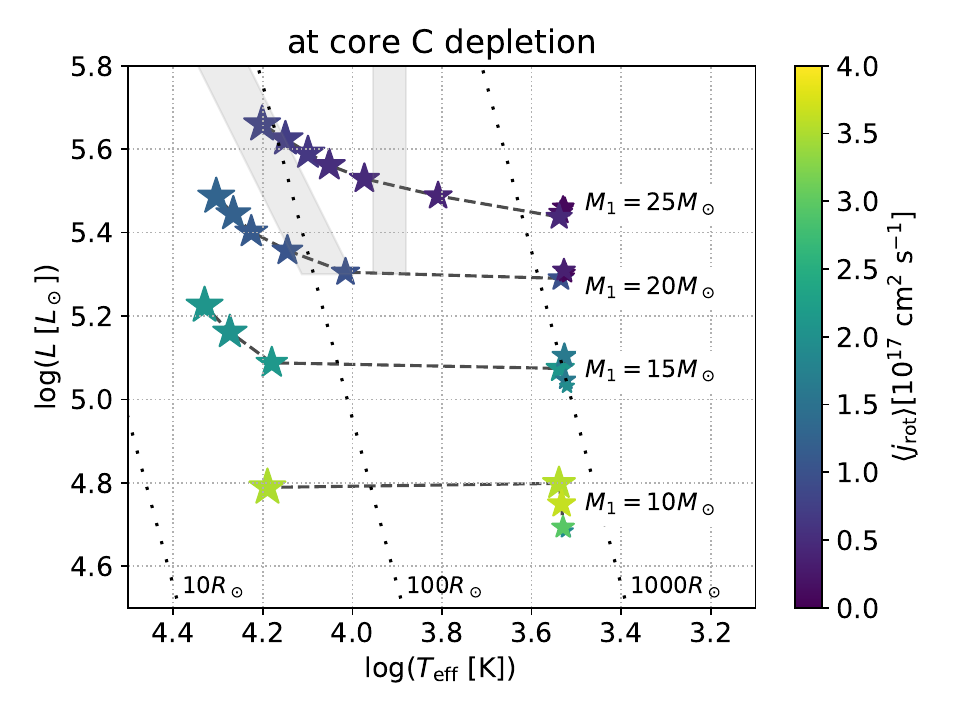}
    \caption{Stellar models on the HR diagram at the time of core carbon depletion. Sizes are larger for higher mass ratio $M_2/M_1$, and colors denote the mass-averaged AM (equation \ref{eq:j_avrg}). Grey shaded regions in the top left show the empirical luminous blue variable (LBV) instability region \citep{Smith04}.}
    \label{fig:HRdiagram_all}
\end{figure}
Comparing the BSG progenitors across different $M_1$, we find a lower specific AM for progenitors with higher $M_1$. This is shown in Figure \ref{fig:HRdiagram_all} with the stellar models color-coded by the mass-averaged specific AM, 
\begin{eqnarray}
\langle j_{\rm rot}\rangle =\frac{1}{M_*}\int^{M_*}_0 j(M_r)dM_r,   
\label{eq:j_avrg}
\end{eqnarray}
that varies by a factor of several across our adopted range of $M_1$, from $3.5\times 10^{17}\ {\rm cm^2\ s^{-1}}$ for the BSG with $M_1=10~M_\odot$ to ($0.5$--$0.8)\times 10^{17}\ {\rm cm^2\ s^{-1}}$ for the models with $M_1=25~M_\odot$ (Table \ref{tab:summary}). We believe this arises mainly due to mass loss after these progenitors contract to BSGs. For a rigidly rotating envelope with gyration radius $kR_*$ and angular velocity $\Omega$, mass loss affects the rotation as
\begin{eqnarray}
    \frac{d(\langle j_{\rm rot}\rangle M_*)}{dt} = -\frac{dM_*}{dt}\left(\frac{2}{3}\Omega R_*^2\right).
    \label{eq:dJdt}
\end{eqnarray}
Using $\langle j_{\rm rot}\rangle=\Omega(kR_*)^2$, we obtain
\begin{eqnarray}
    \frac{d\langle j_{\rm rot}\rangle/dt}{\langle j_{\rm rot}\rangle} = \left(\frac{1}{k^2}-\frac{2}{3}\right)\frac{dM_*/dt}{M_*} ,
\end{eqnarray}
which can be solved, for initial values $\langle j_{\rm rot}\rangle=j_0, M_*=M_{*,0}$, as
\begin{eqnarray}
    \langle j_{\rm rot} \rangle = j_0 \left(\frac{M_*}{M_{*,0}}\right)^{\!\!(1/k^2-2/3)}.
    \label{eq:j_of_M}
\end{eqnarray}
Inspecting our BSG models (solid lines in Figure \ref{fig:jrot_all}), the fractional mass lost during the BSG phase is similar among the progenitor models with a range $5$--$10~\%$. The key difference is the gyration factor $k$ measured in our models from $k^2=\langle j_{\rm rot}\rangle/(\Omega R_*^2)$, ranging from $k^2\approx 0.03$ for the BSG model of $M_1=10M_\odot$ to $k^2\approx 0.01$ for BSGs of $M_1=25M_\odot$. As $k^2\ll 1$, this results in a huge difference in the power-law index in equation \ref{eq:j_of_M}. While the above framework is approximate, this qualitatively explains the much slower rotation obtained for the stellar models with higher $M_1$. This variation with $M_1$ is also seen in models that end as RSGs (dotted lines in Figure \ref{fig:jrot_all}), but with a stronger dependence aided by the higher mass loss rates of RSG winds compared to BSG winds. 

The grey shaded region on Figure \ref{fig:jrot_all} is the threshold AM for a shell to circularize around a BH of mass $M$ and dimensionless spin $a_{\rm BH}=0$--$1$ \citep{Fuller22},
\begin{eqnarray}
    j_{\rm ISCO}(M) &=& \frac{2}{3\sqrt{3}}\left(1+2\sqrt{3r'_{\rm ISCO}-2}\right)\frac{GM}{c} \nonumber \\
    &\sim& 10^{17}\ {\rm cm^2\ s^{-1}}\ \frac{1+2\sqrt{3r'_{\rm ISCO}-2}}{3\sqrt{3}}\left(\frac{M}{15\ M_\odot}\right),
    \label{eq:j_ISCO}
\end{eqnarray}
where $c$ is the speed of light, and $r'_{\rm ISCO}\equiv r_{\rm ISCO}/(GM/c^2)$ ($=1$--$6$) is the innermost stable circular orbit of a BH scaled by $GM/c^2$,
\begin{eqnarray}
    r'_{\rm ISCO} &=& 3+z_2-\sqrt{(3-z_1)(3+z_1+2z_2)}, \nonumber \\
    z_1 &=& 1+(1-a_{\rm BH}^2)^{1/3}\left[(1+a_{\rm BH})^{1/3} + (1-a_{\rm BH})^{1/3}\right], \nonumber \\
    z_2 &=& \sqrt{3a_{\rm BH}^2 + z_1^2}.
\end{eqnarray}
As we discuss in Section \ref{sec:transients}, comparison with $j_{\rm ISCO}$ is most relevant for cases where the progenitor fails to explode upon core-collapse and forms a BH. This $j_{\rm ISCO}$ serves as a benchmark separating the inner part of the star that falls into the newborn BH almost radially, and the outer part that can circularize around the BH to form an accretion disk. We find that the variations in the AM profile for different $M_1$ could result in huge changes in the amount of mass that can circularize around the newborn BH. For example, for the $M_1=10~M_\odot$ model around half of the star's mass can have specific AM exceeding $j_{\rm ISCO}$, while for the $M_1=25~M_\odot$ models only the outermost $0.1$--$1~M_\odot$ of the star can have such large specific AM (see also Table \ref{tab:summary}). Such difference in mass have important consequences in the resulting high-energy transients, as we discuss in Section \ref{sec:failed_explosion}. 

Since the mass loss affects the final AM, the degree of rotation depends on the (unfortunately uncertain) wind prescription. We check this by running additional calculations for the fiducial $15M_\odot+8M_\odot$ BSG model, with wind efficiency factor varied by factor 2 on both sides (i.e. \texttt{Dutch\_scaling\_factor} = 0.25, 1). The average AM $\langle j_{\rm rot}\rangle$ is found to mildly vary by $\approx 30\%$ on both sides (Appendix \ref{app:model_variations}), and other surface properties vary only by a few \%. We thus expect our qualitative picture of the model grid to still hold.

This additionally indicates a  metallicity dependence of our finding, where lower metallicities leading to weaker winds would be more favorable to produce rapidly rotating BSGs. However, it is also noteworthy from our models that the stellar merger channel does not require extremely low metallicities to produce fast-rotating BSGs. Previous stellar modeling, mainly in the context of ultra-long GRBs, have adopted much lower metallicities than ours when producing rotating BSGs ($Z=0$--$2\times 10^{-4}$; \citealt{Kashiyama13,Perna18,Ror24}). Such extremely low metallicities, which are rather unrealistic in the redshifts these events are observed \citep[e.g., Figure 6 of][]{Neijssel19}, are not a requirement for the stellar merger channel.

\section{Transients upon Core-collapse}
\label{sec:transients}
Here we consider the expected transients from the core-collapse of these RSGs and BSGs from merger products. Besides the diversity due to the progenitors, we further expect a more crucial dichotomy in the transient signatures from whether the neutrino-driven explosion at the inner part of the star is successful or not. We respectively discuss the two cases where the neutrino heating drives a successful explosion (Section \ref{sec:successful_explosion}), and where the neutrino-driven explosion fails and a BH forms in the center (Section \ref{sec:failed_explosion}).

As a whole, we expect most of our stellar models to successfully explode by neutrino heating, while some of them fail to explode by neutrinos. Which stars do one or the other is more difficult to predict, as it requires core-collapse SN simulations with accurate initial conditions for the innermost core structure. Absent of first-principle simulations for neutrino-driven explosions of merger products, we here take a model-agnostic approach and include all of our progenitors as candidates for either a successful or failed explosion. Future numerical works with the aforementioned capabilities would be able to narrow down the transient landscape further (see also discussion in Section \ref{sec:conclusion}).

\subsection{The Case of Successful Neutrino-driven Explosions: From SN II-P to 1987A-like SNe}
\label{sec:successful_explosion}

\begin{figure*}
   \centering
   \includegraphics[width=\linewidth]{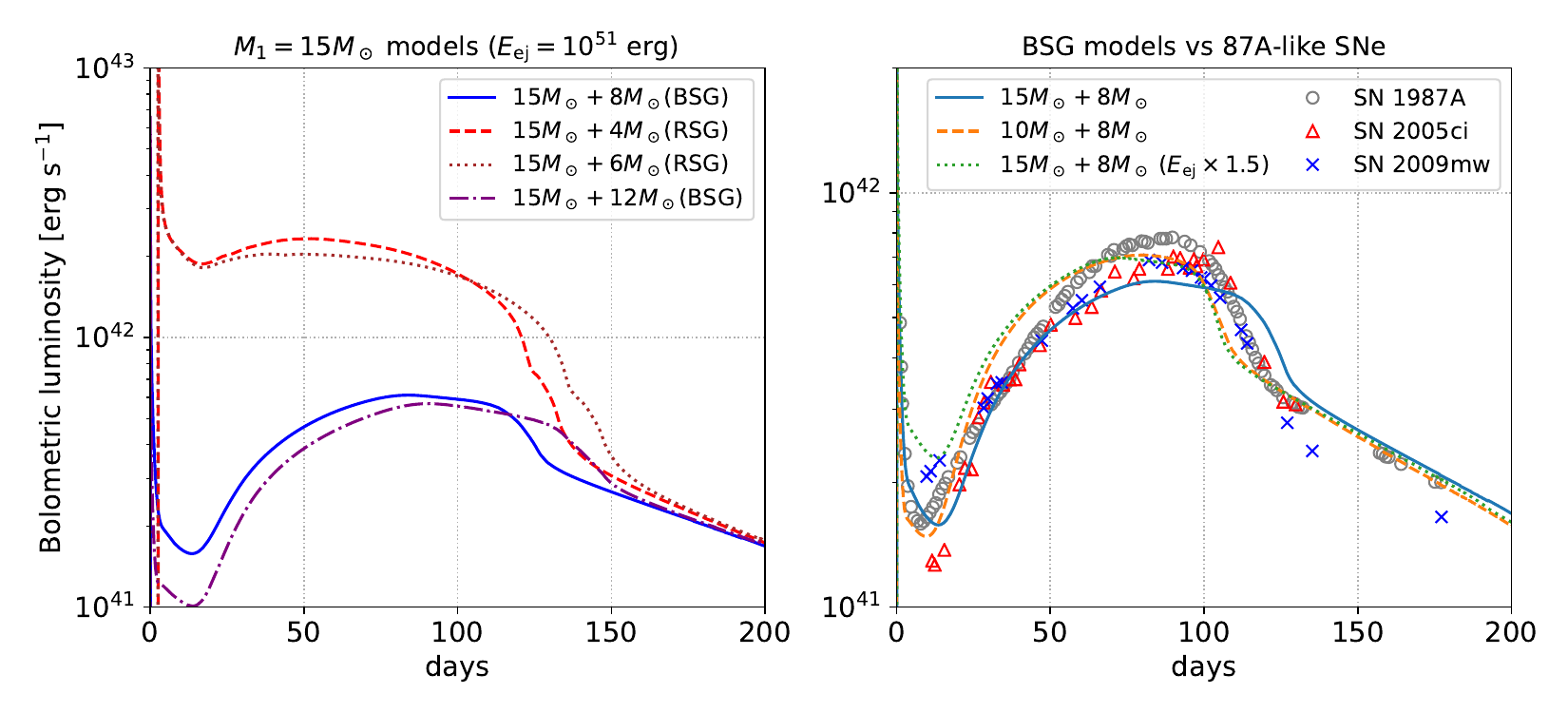}
    \caption{Bolometric light curves for explosions of our merger models from SNEC \citep{Morozova15}. Left panel shows stellar models with $M_1=15~M_\odot$ and varying $M_2$, fixing $E_{\rm ej}=10^{51}$ erg. Right panel shows explosions of representative BSG models, also with increaesd $E_{\rm ej}=1.5\times 10^{51}$ erg for the $15M_\odot+8M_\odot$ model. Markers show the bolometric light curves of SN 1987A \citep{Hamuy88}, SN 2005ci \citep{Taddia16}, and SN 2009mw \citep{Takats16}.}
    \label{fig:BSG-SN}
\end{figure*}

\subsubsection{Supernova Light Curves}
It is well known that the light curves from explosions of RSGs and BSGs greatly differ, due to differences in their radii. The former, typically a Type II-P SNe, shows a characteristic plateau in the light curve powered by diffusion of internal energy deposited by the SN shock \citep{Litvinova83,Popov93,Kasen09}. In the case of more compact BSG progenitors, the internal energy from the SN shock is quickly lost by adiabatic expansion, and the main powering source of the SN is the radioactive decay of $^{56}$Ni and $^{56}$Co. The interplay of radioactive heating and the gradual recombination of the ejecta creates a light curve rising slowly over months, as seen in SN 1987A and reproduced in a number of light curve models \citep[e.g.,][]{Arnett89,Shigeyama90,Utrobin93,Blinnikov00,Moriya24,Pumo25}.

To demonstrate this for our stellar models, we calculate bolometric light curves from explosions of our pre-SN models using a 1D radiation hydrodynamical code SNEC \citep{Morozova15}. Assuming spherical symmetry as initial conditions is a good approximation even for our rotating models, as the stars are rotating at speeds much lower than their mass-shedding limits, and centrifugal forces that deform the star are weak. While multi-D models have demonstrated that the early emission around shock breakout is affected by asymmetries in the near-surface stellar structure or in the explosions themselves \citep{Goldberg22,Irwin21,Vartanyan25}, we expect the later emission at the main peak or plateau to be more robust to the effects of asymmetry.

For all the simulated explosions, we excise the innermost $1.4~M_\odot$ of the pre-SN models assuming it becomes a neutron star. We then adopt SNEC's thermal bomb formulation to inject internal energy over 1 sec in the innermost $0.1~M_\odot$ of the leftover star, with energy set such that the final energy of the ejecta becomes a user-specified value $E_{\rm ej}$. The computational region is sampled by 3000 Lagrangian cells, with finer sampling near the center and the surface.

We fix the mass of $^{56}$Ni in the ejecta as $0.07~M_\odot$, a value inferred in SN 1987A from the $^{56}$Co-powered tail. The degree of mixing of $^{56}$Ni in the ejecta affects the light curve morphology and duration, particularly for BSG explosions. While mixing itself is a multi-D phenomenon, some 1D codes like SNEC allow users to parameterize mixing of $^{56}$Ni by hand. We set $^{56}$Ni to be efficiently mixed in the ejecta out to $90\%$ in mass, motivated by 3D simulations on core-collapse SNe from both BSGs and RSGs \citep[e.g.,][]{Hammer10,Utrobin15,Stockinger20,Vartanyan25}.

The left panel of Figure \ref{fig:BSG-SN} compares the explosions of four merger models with $M_1=15~M_\odot$ and $E_{\rm ej}=10^{51}$ erg, with two stars ($M_2=4,6~M_\odot$) being RSGs and the other two stars ($M_2=8,12~M_\odot$) being BSGs. There is a clear difference in the light curves of these two kinds of explosions over the first $\sim 100$ days, with a IIP-like morphology in the two RSG models and a slowly rising SN 1987A-like morphology for the two BSG models. For a fixed $E_{\rm ej}$, the more massive merger products produce a slightly dimmer light curve with longer duration, due to the larger ejecta mass resulting in a larger photon diffusion time.

The right panel of Figure \ref{fig:BSG-SN} shows three BSG explosion models with varied progenitor and explosion energy, and compared to the bolometric light curves of SN 1987A \citep{Hamuy88}, SN 2005ci \citep{Taddia16}, and SN 2009mw \citep{Takats16}, which all have comparable luminosities at tail such that the required mass of $^{56}$Ni are similar. While variations in the aforementioned parameters slightly change the onset and morphology of the main peak, the overall light curve evolution of (i) a fast initial drop from shock cooling emission of the outermost ejecta, (ii) a slow rise to main peak, and (iii) fast fall to the $^{56}$Co-powered tail due to hydrogen recombination, is in good qualitative agreement to the observed events. An additional parameter not captured in the SNEC calculations is the mixing of hydrogen in the ejecta, which can affect the duration of the main peak governed by hydrogen recombination. Modeling of light curve and nebular spectra for SN 1987A imply some mixing of hydrogen in the inner core \citep{Shigeyama90,Utrobin93,Kozma98}, but detailed tuning to match the light curve of SN 1987A is beyond the scope of this work (see e.g. \citealt{Utrobin15,Menon19} for such attempts).

\subsubsection{Event rates of 1987A-like SNe}
\label{sec:1987A_rates}
The rate of 1987A-like SNe from BSG progenitors is suggested to be 1--3 \% of core-collapse SNe \citep{Smartt09,Kleiser11,Pastorello12,Taddia16,Sit23}, with a preference for sub-solar metallicities comparable to LMC \citep{Taddia13,Sit23}. As we cover the mass range of progenitors where we expect a large fraction of them to explode, the observed rates is a good cross check of our binary models. To estimate the event rate of these SNe from our binary channel, we conduct a similar exercise to \cite{Justham14}.

From the results in Section \ref{sec:model_grid}, we infer that binaries with $q\gtrsim 0.5$--$0.6$ end up as BSGs upon core-collapse. Our MESA models do not constrain the initial separation of the binaries required to produce BSGs, but this can be roughly predicted from binary evolution. At large separations of $a\gtrsim$ (a few) $10^3~R_\odot$, the two stars do not interact, and evolve as single stars that become RSGs at core-collapse (or Wolf-Rayet stars for ZAMS mass $\gtrsim 35~M_\odot$; \citealt{Sukhbold16}). At intermediate separations of $a\approx$ (a few) $100$-- (a few) $1000~R_\odot$ the binary can interact as the primary crosses the Hertzsprung gap, but a clear core-envelope boundary would exist in the primary. The primary's envelope is loosely bound such that it is likely to be ejected in the common envelope phase \citep[e.g.,][]{Klencki21} at such high values of $q$, making the primary a hydrogen-poor star. At the tightest separations of $a\lesssim 10~R_\odot$ the binaries merge earlier in the main sequence \citep[e.g.,][]{deMink13,Kinugawa24}, and rejuvenates as a single main-sequence star that most likely ends its life as a RSG or a Wolf-Rayet star \citep{Schneider24}. We therefore expect (early) Case B mergers leading to BSGs to occur in binaries with initial separations of $a\approx 10$ -- (a few) $100~R_\odot$. For a log-uniform distribution of $a$ spanning 5--6 dex commonly adopted in population synthesis modeling, binaries in this separation range occupy $f_a\sim 20$--$30\%$ of the entire binary population. 

Further adopting a binary fraction of $f_{\rm bin}\sim 50$\% for massive stars in LMC-like environments \citep{Sana13,Dunstall15} and a flat mass ratio distribution \citep{Shenar22} with binaries of $0.5 (0.6)\leq q\leq 0.8$ occupying $f_q\approx 30\% (20\%)$, we obtain the core-collapses of BSGs from our merger channel to be $f_af_{\rm bin}f_q\approx$ $2$--$4\%$ of all core-collapses. Our estimate reasonably agrees with the observed fraction of 1987A-like SNe, given the expectation that a majority of our BSGs explode. This implies that binaries in sub-solar metallicity environments play a key role in producing 1987A-like SNe. 

As the threshold for $q$ to undergo a contact phase is somewhat uncertain, the rate would be reduced if one considers a lower threshold of $q$ for a contact phase to occur. Meanwhile we expect an additional contribution from wider, low-$q$ binaries undergoing Case C mergers (as proposed for SN 1987A), although it requires the separation to be further tuned to a narrow range for unstable Case C mass transfer to occur while avoiding unstable Case B mass transfer.

\subsection{The Case of Failed Neutrino-driven Explosions: Accretion-driven Transients}
\label{sec:failed_explosion}
If the neutrino-driven explosion fails and a BH forms, the fate of the leftover star is completely different from a successful SNe. The loss of pressure support in the center would lead to infall of the inner part of the star onto the newborn BH, while mass loss due to neutrino emission preceding BH formation could eject the outermost part of the star \citep{Nadyozhin80,Lovegrove13,Fernandez18,Ivanov21}. In our progenitors with rotating envelopes, we may expect an energetic transient powered by accretion of stellar material onto the BH \citep[e.g.,][]{Dexter13,Kashiyama15,Moriya18}. 

For RSGs however, we expect that a large fraction (or all) of the outer envelope is ejected due to neutrino mass loss \citep{Lovegrove13,Fernandez18,Ivanov21}. Even if the inner envelope could fall onto the BH, the AM from rotation ($\lesssim 10^{17}$--$10^{18}$ cm$^2$ s$^{-1}$; Figure \ref{fig:jrot_all}) is subdominant with respect to that expected from random convective motion ($10^{18}$--$10^{19}$ cm$^2$ s$^{-1}$; \citealt{Goldberg22a,Antoni22}). Therefore coherent circularization of the envelope in the form of an accretion disk is likely not realized for RSG progenitors, and it has instead been suggested that the randomly-oriented infall will unbind the envelope via a weak explosion of energies $10^{48}$--$10^{49}$ ergs \citep{Antoni23}. The resulting transient has been investigated in past works, and is expected to have a shock breakout emission of luminosity $10^7$--$10^8L_\odot$ that lasts for days to weeks \citep{Piro13,Lovegrove17,Tsuna25}, followed by a months to year-long plateau emission with luminosity $10^6$--$10^7L_\odot$ \citep{Lovegrove13,Antoni23}.

On the other hand, BSGs have compact envelopes that results in weak mass ejection ($<0.1~M_\odot$) by the neutrino mass loss \citep{Fernandez18,Ivanov21}. As the envelope is radiative and rotation dominates its AM content, coherent circularization of the envelope would be realized. The envelope falls back over the free-fall time at the stellar surface
\begin{equation}
    t_{\rm ff}\approx \sqrt{\frac{\pi^2 R_*^3}{8GM_{\rm *}}} \sim 1\ {\rm day}\left(\frac{R_*}{50R_\odot}\right)^{\!\!3/2}\left(\frac{M_{\rm *}}{30M_\odot}\right)^{\!\!-1/2}.
    \label{eq:t_freefall}
 \end{equation}
We find in Figure \ref{fig:jrot_all} that the material able to circularize outside the BH has masses of $M_{\rm circ}\sim 0.1$--$10~M_\odot$, local free-fall times of $10^3$--$10^5$ sec, and bulk specific AM of $j\sim 10^{17}$--$10^{18}$ cm$^2$ s$^{-1}$. The circularization of fallback matter forms an accretion disk at a characteristic radius
\begin{eqnarray}
    r_{\rm disk} \approx \frac{j^2}{GM_{\rm BH}}\sim 4\times 10^8\ {\rm cm} \left(\frac{j}{10^{18}\ {\rm cm^2\ s^{-1}}}\right)^{\!\!2} \left(\frac{M_{\rm BH}}{20\ M_\odot}\right)^{\!\!-1},
\end{eqnarray}
and accretes onto the BH at a viscous timescale of
\begin{eqnarray}
    t_{\rm visc} &\approx& \frac{1}{\alpha(H/R)^2}\sqrt{\frac{r_{\rm disk}^3}{GM_{\rm BH}}} \nonumber \\
    &\sim& 4\ {\rm sec}\left[\frac{\alpha(H/R)^2}{0.03}\right]^{-1} \left(\frac{j}{10^{18}\ {\rm cm^2\ s^{-1}}}\right)^{\!\!3} \left(\frac{M_{\rm BH}}{20\ M_\odot}\right)^{\!\!-2},
\end{eqnarray}
where $\alpha$ is the viscosity parameter \citep{Shakura73}, and $H/R$ is the height-to-radius ratio. As $t_{\rm visc}\ll t_{\rm ff}$, the accretion rate is regulated not by viscosity but by the infall of the envelope, with accretion rates of $\dot{M}_{\rm acc}\sim M_{\rm circ}/t_{\rm ff}\sim 10^{-5}$--$10^{-3}\ M_\odot\ {\rm s^{-1}}$. This is orders of magnitude higher than the Eddington-limited accretion rate where photons can cool the accretion flow, yet lower than the rate where neutrinos can cool the flow ($\gtrsim 10^{-2}\ M_\odot\ {\rm s}^{-1}$; \citealt{Chen07,Kawanaka13}). For such radiatively inefficient accretion flows, we expect outflows from the disk \citep{Shakura73,Blandford99}, with high velocities of the order of the escape velocity within the disk $v_{\rm wind}\gtrsim \sqrt{GM_{\rm BH}/r_{\rm disk}}\sim 0.1c$ \citep[see also][]{Strubbe09,Kashiyama15}.

A subset of the BSG models with the largest accretion rates may further launch relativistic jets, analogous to the collapsar scenario for long GRBs \citep{Macfadyen99}. Such ``BSG collapsars" have previously been proposed as channels for ultra-long GRBs \citep[e.g.,][]{Woosley12,Gendre13,Kashiyama13,Nakauchi13,Perna18}, as they can naturally realize engine durations ($\sim t_{\rm ff}$) consistent with the observed durations of these GRBs. While there are only a handful of observed ultra-long GRBs, one event (GRB 111209A) has a superluminous SN-like optical counterpart (SN 2011kl), potentially consistent with this scenario \citep{Levan14,Greiner15}.

\subsubsection{BSG Collapsars: Envelope Accretion}
\label{sec:collapsar_model}

To investigate the expected transients in detail, we model the time-dependent accretion of BSG collapsars using the model of \cite{Fuller22} (their Appendix A). The disk is evolved by a one-zone approach that includes infall of stellar material, viscous accretion, and the wind outflow. The disk forms when the specific AM $j$ of the infalling shell of material first exceeds the $j_{\rm ISCO}$ of the BH, for a BH mass and spin given by assuming all the material interior to this shell collapses into a BH. The disk material is then continuously sourced by infall of stellar material with $j>j_{\rm ISCO}$, and is taken away by both accretion onto the BH and the disk wind. 

At any given time the disk is characterized by two parameters, its mass $M_d$ and total AM $J_d$. The viscous evolution of the disk is calculated for $\alpha(H/R)^2=0.03$. In our BSG models where the accretion rate is regulated by the free-fall time of the stellar material rather than the viscous time, the accretion is insensitive to this choice.

Radiatively inefficient accretion flows are expected to be geometrically thick, with aspect ratios of $H/R\sim 0.3$--$1$, and are prone to powerful outflows. The model assumes that once the disk forms, only the equatorial material within a given solid angle of the disk manage to fall back onto the disk, and polar material outside this solid angle is removed by mechanical feedback from the disk wind/jet. We adopt a constant disk opening angle of $\theta_{\rm disk}=45$ degrees, corresponding to $H/R=1$. This results in a fraction $\sin(\theta_{\rm disk})\approx 0.7$ of stellar material with $j>j_{\rm ISCO}$ contributing to the disk. 

The outermost bit of the envelope is expected to be ejected by the aforementioned neutrino mass loss, and will not contribute to the infalling mass. The ejected mass most sensitively depends on the compactness of the envelope $\xi_{\rm env}=(M_*/M_\odot)/(R_*/R_\odot)$ as defined in \cite{Fernandez18}, which is expected since the binding energy of an outermost shell of a given mass $\Delta M(\ll M_*)$ scales with $M_*/R_*$. For each stellar model we adopt an ejected mass inversely scaling with $\xi_{\rm env}$ as $M_{\rm ej}\approx0.01\xi_{\rm env}^{-1}\ M_\odot$ based on the simulations of \cite{Ivanov21}, which results in $M_{\rm ej}\approx 0.01$--$0.1~M_\odot$ for our BSG models.

With these prescriptions, we solve for the time-evolution of the disk parameters using eq (A1)--(A4) of \cite{Fuller22}. The radial mass-accretion rate in the disk down to $r_{\rm ISCO}$ is expected to follow a power-law $\dot{M}(r)=(M_d/t_{\rm visc})(r/r_d)^{s}$ \citep{Blandford99}, where $r$ is the radius from the BH and we set $s=0.5$ motivated from recent simulations of radiatively inefficient accretion flows \citep{Cho24,Guo24}. The time-dependent mass-loss rate and kinetic luminosity of the wind is then given by integration over $r$ as
\begin{eqnarray}
    \dot{M}_{\rm wind}(t) &\approx& \frac{M_d}{t_{\rm visc}} \left[1- \left(\frac{r_{\rm ISCO}}{r_{\rm disk}}\right)^{\!\!s} \right], \\
    L_{\rm wind}(t) &\approx& \frac{s}{2(1-s)}\frac{GM_{\rm BH}}{r_{\rm disk}}\frac{M_d}{t_{\rm visc}} \left[\left(\frac{r_{\rm disk}}{r_{\rm ISCO}}\right)^{\!\!1-s}-1\right].
\end{eqnarray}
Finally the accretion onto the BH is given by
\begin{eqnarray}
    \dot{M}_{\rm BH}(t)=\dot{M}_{\rm acc}- \dot{M}_{\rm wind}= \frac{M_d}{t_{\rm visc}} \left(\frac{r_{\rm ISCO}}{r_{\rm disk}}\right)^{s}.
\end{eqnarray}

\begin{figure}
   \centering
   \includegraphics[width=\linewidth]{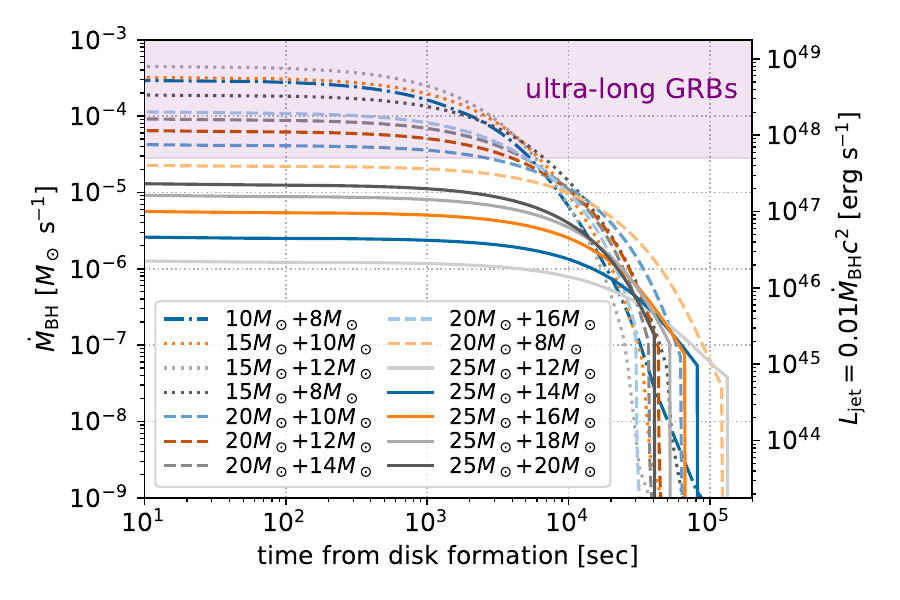}
    \caption{Accretion history of the newborn BH for the BSG progenitors, in the case they underwent a failed SN. Line styles are varied by $M_1$. Right axis shows the corresponding jet power $L_{\rm jet}=\eta_{\rm jet}\dot{M}_{\rm BH}c^2$, for a jet efficiency $\eta_{\rm jet}=0.01$. The shaded region shows the range of jet power inferred from observed ultra-long GRBs (see Section \ref{sec:ulGRBs}).}
    \label{fig:Mdot_BH}
\end{figure}

Figure \ref{fig:Mdot_BH} shows the accretion history onto the BH $\dot{M}_{\rm BH}(t)$ from disk formation, when the first (innermost) infalling material circularizes around the BH's ISCO. We find $\dot{M}_{\rm BH}$ to be most sensitive to the primary mass, as expected from the large differences of the specific AM among these masses. The four $M_1=10,15\ M_\odot$ models (dash-dotted/dotted lines), the five $M_1=20\ M_\odot$ models (dashed lines), and the five $M_1=25\ M_\odot$ models (solid lines) respectively predict accretion rates of $10^{-4}$--$10^{-3}M_\odot\ {\rm s}^{-1}$, $10^{-5}$--$10^{-4}M_\odot\ {\rm s}^{-1}$, and $10^{-6}$--$10^{-5}M_\odot\ {\rm s}^{-1}$. The durations of the high accretion phase are in the range of $1$--$10$ hours, longer for stars with higher $M_1$ due to their larger radii (Table \ref{tab:summary}; equation \ref{eq:t_freefall}).

\subsubsection{BSG Collapsars: Implications for Ultra-long GRBs}
\label{sec:ulGRBs}

\begin{figure}
   \centering
   \includegraphics[width=\linewidth]{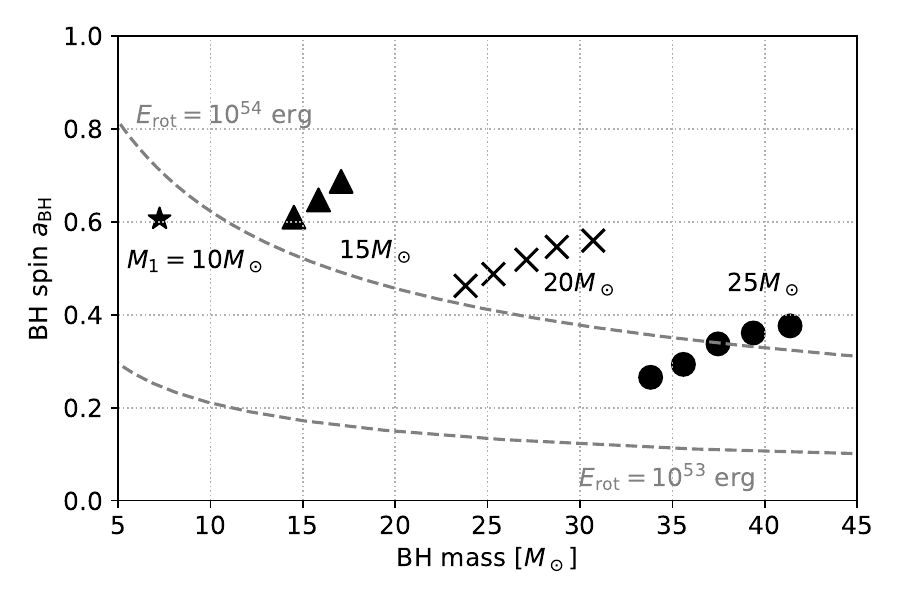}
    \caption{BH mass versus spin at the time of disk formation for the BSG models, with different markers indicating different values of $M_1$. These parameters, particularly the spin, are likely to change by the subsequent disk evolution. Dashed curves show the corresponding BH rotation energies (equation \ref{eq:E_rot}) of $10^{53}$, $10^{54}$ erg.}
    \label{fig:Mbh_vs_abh}
\end{figure}

Ultra-long GRBs are a class of GRBs with very long durations of hours and isotropic-equivalent luminosities of $\sim 10^{49}$--$10^{50}$ erg s$^{-1}$ \citep{Levan14}. The $\gamma/$X-ray emission of the burst has non-thermal spectra and rapid variability similar to classical GRBs, indicating beamed emission from a relativistic jet.

In classical GRBs, the large energies in the relativistic jets have led to suggestions of them being generated electromagnetically \citep[e.g.,][]{Lyutikov03,Kawanaka13,Liu15, kumar15_GRB_jets}, where the jet is powered by the extraction of rotational energy from the BH assisted by large-scale magnetic fields \citep{Blandford77}. Figure \ref{fig:Mbh_vs_abh} shows the BH spin of our BSGs models at the start of disk formation, calculated by the model in Section \ref{sec:collapsar_model}. The BH has moderate spin for all cases with $a_{\rm BH}\approx 0.2$--$0.7$, inherited from the finite AM of stellar material plunging into the BH. The rotational energies that can in principle be extracted from the BHs \citep[e.g., equation 18 of][]{Rees84},
\begin{eqnarray}
    E_{\rm rot}=\left(1-\sqrt{\frac{1+\sqrt{1-a_{\rm BH}^2}}{2}}\right)M_{\rm BH}c^2,
    \label{eq:E_rot}
\end{eqnarray}
are of the order of $10^{54}$ erg, sufficient to power a strong jet with energies observed in ultra-long GRBs.

Recent studies \citep{Jacquemin-Ide24,Lowell24} have considered the BH spin evolution in collapsar disks, under the  scenario that the magnetic field is strong enough such that the BH-disk system could reach the magnetically-arrested disk (MAD) state \citep{Narayan03,Tchekhovskoy11}. In this MAD regime, two key findings are: (i) The jet luminosity scales with the accretion rate with a spin-dependent efficiency $L_{\rm jet}=\eta_{\rm jet}(a_{\rm BH})\dot{M}c^2$, and (ii) The spin of the BH is quickly reduced to $a_{\rm BH}\approx 0.1$--$0.2$ even if only a small fraction of the original mass is accreted, which drops the jet efficiency to $\eta_{\rm jet}\sim 0.4$--$2\%$. More recent work \citep{Wu25} extended this framework to non-MAD disks, finding a maximal jet efficiency of $\eta_{\rm jet}\sim 2\%$ and lesser degree of spindown when the system deviates from MAD.

{\it If we assume that} the efficiencies for long GRBs similarly apply in our case of BSG collapsars (although with a few orders of magnitude lower accretion rate), $L_{\rm jet}$ could be calculated as shown in Figure \ref{fig:Mdot_BH} for $\eta_{\rm jet}=1\%$. The beaming-corrected average X-ray luminosity of $\gtrsim 10^{47} {\rm erg\ s^{-1}}$ inferred in ultra-long GRBs \citep[][]{Levan14}, or a jet power of $\gtrsim 5\times 10^{47} {\rm erg\ s^{-1}}$ for a radiative efficiency of $20\%$ \citep{Beniamini16}, can be achieved for BSG models with $M_1=10,15,20~M_\odot$. These also meet the required duration of hours.

\begin{figure}
   \centering
   \includegraphics[width=\linewidth]{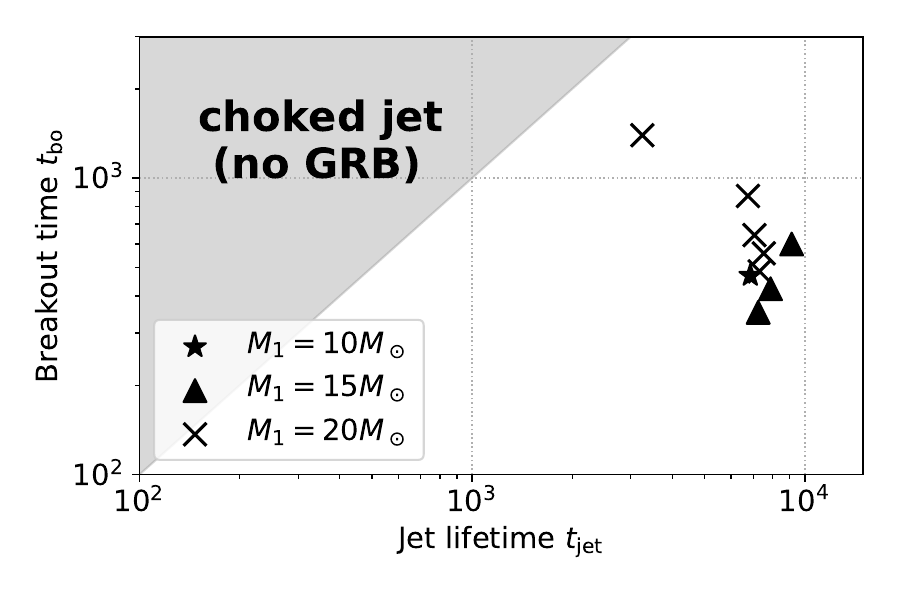}
    \caption{Comparison between the jet duration (set by threshold $L_{\rm jet}>3\times 10^{47}$ erg s$^{-1}$) and the time for jet to break out from the star, for the nine BSG models with $M_1=10,15,20\ M_\odot$. The shaded region of $t_{\rm jet}<t_{\rm bo}$ represents the regime where the relativistic jet is choked in the star, and no prompt emission is expected to be observable.}
    \label{fig:jet_propagation}
\end{figure}

To observe the ultra-long GRB, the relativistic jet formed near the BH has to break out of the BSG envelope before the accretion ceases. To check this constraint, we calculate (see Appendix \ref{app:jet_propagation}) the propagation of a jet inside the star with a model based on \cite{Bromberg11}. Figure \ref{fig:jet_propagation} shows the comparison between the time it takes for a jet to break out the BSG envelope $t_{\rm bo}$, and the lifetime of the jet $t_{\rm jet}$. The latter is set by a threshold on the jet power of $L_{\rm jet}>3\times 10^{47}$ erg s$^{-1}$, but the dependence of $t_{\rm jet}$ on this is weak because of the sharp drop in accretion at late times. For all models we find $t_{\rm jet}>t_{\rm bo}$, i.e. the jet breaks out the star before it shuts off, and emission from the jet is observable for a duration $\approx t_{\rm jet}-t_{\rm bo}\sim 10^3$--$10^4$ sec.

While the jet propagates through the star, a fraction $\sim t_{\rm bo}/t_{\rm jet}\sim 10\%$ of the jet energy is deposited into the cocoon. We expect thermal emission from the nearly isotropic (outer) cocoon, especially visible for off-axis observers \citep[e.g.,][]{Zheng25}. The cocoon's mass from our jet model is found to range as $M_c\approx0.1$--$0.6~M_\odot$, with asymptotic velocities $v_c\approx \sqrt{2E_c/M_c}$ approaching $\sim 0.1c$. 
The timescale of the outer cocoon emission is set by its diffusion time
\begin{eqnarray}
    t_{\rm oc}\approx\sqrt{\frac{\kappa M_c}{4\pi v_c c}} 
    \sim 4\ {\rm day} \left(\frac{v_{\rm c}}{0.1c}\right)^{\!\!-1/2}\left(\frac{M_{\rm c}}{0.2~M_\odot}\right)^{\!\!1/2},
\end{eqnarray}
where we adopted a uniform opacity of $\kappa\approx 0.3\ {\rm cm^2\ g^{-1}}$. The cocoon adiabatically expands from an initial volume at breakout of $V_{c,0}=f_c R_*^3$, where $f_c\approx 0.08$--$0.13$ in our models, and radiates its remaining internal energy over a timescale of $t_{\rm oc}$ when it expands to a volume of $\sim 4\pi(t_{\rm oc}v_c)^3/3 \gg V_{c,0}$. This leads to a luminosity
\begin{eqnarray}
    L_{\rm oc}&\approx& \frac{E_c}{t_{\rm oc}}\left[\frac{f_cR_*^3}{4\pi (v_ct_{\rm oc})^3/3}\right]^{1/3} \nonumber \\
    &\approx&\frac{(6\pi^2f_c)^{1/3}R_*v_c^2c}{\kappa} \nonumber \\
    &\sim& 6\times 10^{42}\ {\rm erg\ s^{-1}} \left(\frac{f_{\rm c}}{0.1}\right)^{\!\!1/3} \left(\frac{R_*}{50~R_\odot}\right)\left(\frac{v_{\rm c}}{0.1c}\right)^{\!\!2},
\end{eqnarray}
and effective temperature
\begin{eqnarray}
  T_{\rm eff, oc} &\approx& \left[\frac{L_{\rm oc}}{4\pi (v_ct_{\rm oc})^2\sigma_{\rm SB}}\right]^{1/4} \nonumber \\
  &\sim& 10000\ {\rm K} \left(\frac{f_{\rm c}}{0.1}\right)^{\!\!1/12} \left(\frac{R_*}{50~R_\odot}\right)^{\!\!1/4}\nonumber \\
  &&\times \left(\frac{M_{\rm c}}{0.2~M_\odot}\right)^{\!\!-1/4}\left(\frac{v_{\rm c}}{0.1c}\right)^{\!\!1/4},
\end{eqnarray}
where $\sigma_{\rm SB}$ is the Stefan-Boltzmann constant. We thus predict an emission peaking in near-UV with AB magnitude of $\approx$ -17 mag at a wavelength of 250 nm, detectable out to $z\approx 0.2$ with ULTRASAT for a single-visit depth of 22.5 mag \citep{Shvartzvald24}.

A final important constraint comes from the event rates of these ultra-long GRBs. While the number of detected ultra-long GRBs are much smaller than the classical long GRBs, ultra-long GRBs have lower luminosities and is visible on a smaller horizon. Taking this into account, \cite{Levan14} inferred the event rate of ultra-long GRBs to be comparable or down to a factor several lower than long GRBs. The local long GRB rate is estimated to be $\sim 1\ {\rm Gpc}^{-3}\ {\rm yr}^{-1}$ \citep{Wanderman10,Lien14}, i.e. $\sim 0.1\%$ of all core-collapse SNe for a beaming fraction of $10^{-2}$, while the core-collapse of BSGs are estimated in Section \ref{sec:1987A_rates} to be (a few) \% of core-collapse SNe. We thus expect at most a few \% of our BSG core-collapses will actually produce the ultra-long GRBs modeled here, regardless of whether the BSG collapsars are the progenitors of ultra-long GRBs \citep[see e.g.][for alternative scenarios]{Macleod14,Ioka16,Marchant20,Hutchinson-Smith24}. This upper limit is rather constraining, given that the rate of failed SNe could be as high as $\sim 20\%$ of massive star core collapse \citep{Neustadt21}. This may indicate that a failed explosion and BH formation have a ZAMS mass distribution biased towards more massive BSGs, that would be less likely to have a massive rotating envelope to realize the extreme accretion.

\subsubsection{BSG Collapsars: Wind-driven Luminous Transients}
\begin{figure}
   \centering  
   \includegraphics[width=\linewidth]{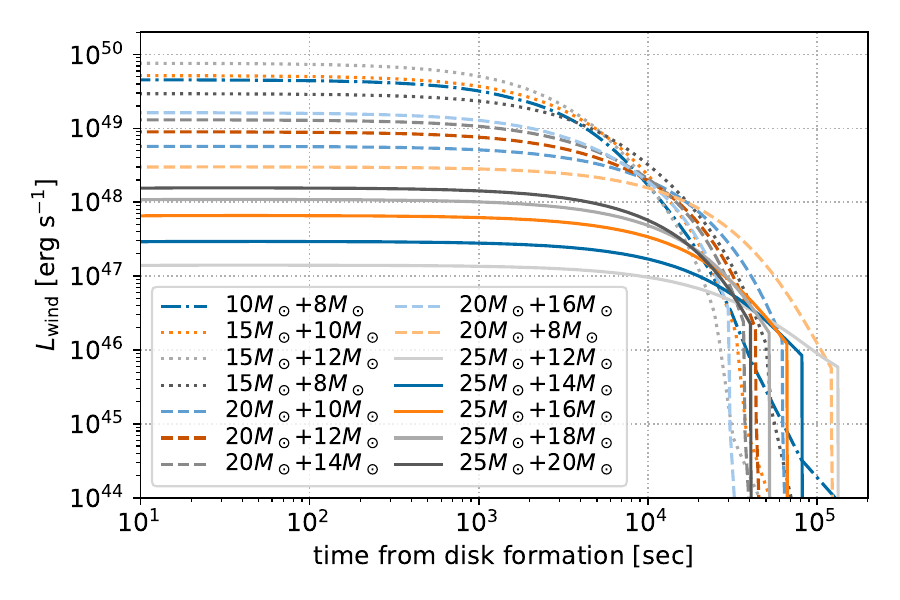}
    \caption{Kinetic luminosity of the disk wind for the same BSG progenitors as in Figure \ref{fig:Mdot_BH}.}
    \label{fig:Lwind}
\end{figure}

Regardless of whether a relativistic jet is launched, the mildly relativistic wind of $v_w\sim 0.1$--$0.3c$ from super-Eddington accretion could also power a bright thermal transient on its own. The wind will eject most of the equatorial disk because $r_{\rm disk}\gg r_{\rm ISCO}$, and the polar part of the star will also be unbound. 

Figure \ref{fig:Lwind} shows the kinetic luminosity of the disk wind $L_{\rm wind}$. The time-integrated energy of the wind varies with $4\times 10^{51}$ -- $2\times 10^{53}$ erg, with the largest energies achieved by the $M_1=10,15~M_\odot$ models ejecting $M_{\rm ej}\sim 10~M_\odot$ of the star, and the least energetic being the $M_1=25~M_\odot$ models that eject $M_{\rm ej}\sim 0.2$--$0.7~M_\odot$.

We estimate the radiative output of this explosion below. As the wind breaks out of the star and the resulting ejecta expands, thermal emission is radiated over a diffusion timescale of 
\begin{eqnarray}\label{eq:tdiff}
    t_{\rm diff}&\approx& \sqrt{\frac{\kappa M_{\rm ej}}{4\pi v_{\rm ej}c}}\nonumber \\
    &\sim& \begin{cases}
       30\ {\rm day}\left(\frac{v_{\rm ej}}{0.1c}\right)^{\!-1/2}\left(\frac{M_{\rm ej}}{10~M_\odot}\right)^{\!1/2}  & (M_1=10,15M_\odot)\\
       6\ {\rm day}\left(\frac{v_{\rm ej}}{0.1c}\right)^{\!-1/2}\left(\frac{M_{\rm ej}}{0.5~M_\odot}\right)^{\!1/2} & (M_1=25M_\odot),
    \end{cases},
\end{eqnarray}
where we have again used $\kappa\approx 0.3\ {\rm cm^2\ g^{-1}}$. As $t_{\rm diff}$ is longer than the duration of the disk wind $t_{\rm acc}\sim 10^3$--$10^4$ s, the ejecta radiates after a phase of adiabatic expansion, the energy supplied by the wind that occurs at a characteristic radius 
\begin{eqnarray}
r_{\rm inj}\approx {\rm max}(R_*, v_{\rm ej}t_{\rm acc}).     
\end{eqnarray}
The former case of $r_{\rm inj}\approx R_*$ is valid for instantaneous energy injection (e.g. core-collapse SNe), while the latter case of $r_{\rm inj}\approx v_{\rm ej}t_{\rm acc}$ is valid for long-lasting energy injection (e.g. magnetar engine; \citealt{Kasen10}). Our case is marginally the latter, where $v_{\rm ej}t_{\rm acc}/R_*$ ranges from unity to a few. Since internal energy drops with radius as $\propto r^{-1}$ by adiabatic expansion, the resulting luminosity is
\begin{eqnarray}
    L&\approx& \left(\frac{r_{\rm inj}}{v_{\rm ej}t_{\rm diff}}\right) \frac{0.5M_{\rm ej}v_{\rm ej}^2}{t_{\rm diff}}\approx \frac{2\pi v_{\rm ej}^{\!3} t_{\rm acc}c}{\kappa} \nonumber \\
    &\sim& \begin{cases}
       2\times 10^{43}\ {\rm erg\ s^{-1}}\left(\frac{v_{\rm ej}}{0.1c}\right)^{\!3}\left(\frac{t_{\rm acc}}{10^3{\rm s}}\right)  & (M_1=10,15M_\odot)\\
       2\times 10^{44}\ {\rm erg\ s^{-1}}\left(\frac{v_{\rm ej}}{0.1c}\right)^{3}\left(\frac{t_{\rm acc}}{10^4{\rm s}}\right)  & (M_1=25M_\odot).
    \end{cases}
\end{eqnarray}
The timescale, luminosity, and the photospheric velocity ($\sim v_{\rm ej}$) predicted from the low-$M_1$ models align with the ``super-luminous SN" component seen in an ultra-long GRB 111209A (SN 2011kl; \citealt{Greiner15}). This matches well with the picture raised in the previous section, of these low-mass BSGs also being more favorable progenitors for ultra-long GRBs.

On the other hand, the $M_1=25~M_\odot$ models predict a $\lesssim$ week-long emission with luminosity reaching $\sim 10^{44}$ erg s$^{-1}$, overlapping with the parameter space of AT2018cow-like fast blue optical transients \citep[FBOTs;][]{Ho23}. As initially proposed for AT2018cow \citep[e.g.,][]{Margutti19}, the BSG collapsar model could explain many of the key observational features of FBOTs, such as the presence of hydrogen and helium, fast and asymmetric ejecta, low $^{56}$Ni yield, and the star-forming host environment.

Still, the current model faces two potential challenges when interpreting the multi-wavelength phenomenology of FBOTs, which merit further investigations. First is the X-ray emission lasting for up to years \citep[e.g.,][]{Coppejans20,Yao22,Migliori24}, which suggests long-lived accretion onto the remnant compact object. While our model does not predict strong accretion over such timescales (Figure \ref{fig:Mdot_BH}), a key component not included in our model is fallback material of the neutrino-driven weak explosion \citep{Fernandez18}. This can create a power-law tail in the accretion rate, that can keep the accretion super-Eddington for years after core-collapse. This material is guaranteed to have a large AM of $10^{18}$--$10^{19}\ {\rm cm^2\ s^{-1}}$, enough to sustain an accretion disk.

Second is the existence of fairly dense circumstellar matter inferred from radio observations \citep[e.g.,][]{Ho19,Margutti19,Coppejans20,Ho20,Ho22,Yao22}.
The inferred mass-loss rates at $\sim 10^{17}$ cm are $\sim 10^{-5}$--$10^{-4}~M_\odot\ {\rm yr}^{-1}$, when scaled for a BSG-like wind velocity of $100$ km s$^{-1}$. The mass-loss rates predicted by the prescriptions in our MESA models at the last centuries of evolution are typically $\sim 10^{-6}M_\odot$ yr$^{-1}$, at least an order of magnitude lower. One possibility to resolve this discrepancy is that the mass loss is underestimated, as the default Dutch wind prescription does not include LBV-like mass loss. Inspecting the stellar tracks of our $M_1=25M_\odot$ BSG models, we find that they typically enter the empirically-known LBV instability region \citep[][see Figure \ref{fig:HRdiagram_all}]{Smith04} when they slightly expand at helium shell burning. The measured mass loss rates for LBV winds is $\sim 10^{-5}$--$10^{-4}~M_\odot\ {\rm yr}^{-1}$ \citep{Smith14}, which matches with the values inferred from radio. However, such extreme mass-loss over the final $\sim 10^3$--$10^4$ yrs of the star from helium shell burning to core-collapse can be problematic, as it could entirely remove the rotating outermost envelope of the star, required for triggering accretion that powers the FBOTs themselves.

\section{Conclusion}
\label{sec:conclusion}

We have presented rotating supergiant models from post-main sequence stellar mergers, using the stellar evolution code MESA with updated prescriptions for AM transport inside the star. In our model, the key parameter that set the dichotomy of a RSG or BSG at death was the mass accreted onto the primary, with larger binary mass ratios (typically $\gtrsim 0.5$--$0.6$) more likely leading to BSGs. Though the merger product loses some mass and AM due to centrifugal mass loss, our BSG models tend to maintain fast rotation in the outer envelope until core-collapse, and models with smaller ZAMS masses lead to faster rotation due to the effects of stellar winds during the helium burning phase and beyond. 

Using the models evolved until close to core-collapse we made predictions for the transients of these stars upon core-collapse, considering both cases where the canonical neutrino-driven explosion succeeds and fails. For the former case of canonical SNe, light curve modeling successfully reproduced the distinct SN light curves from RSGs and BSGs, i.e. the Type II-P and the Type II-pec (or 1987A-like) SNe.

\begin{figure*}
   \centering  \includegraphics[width=\linewidth]{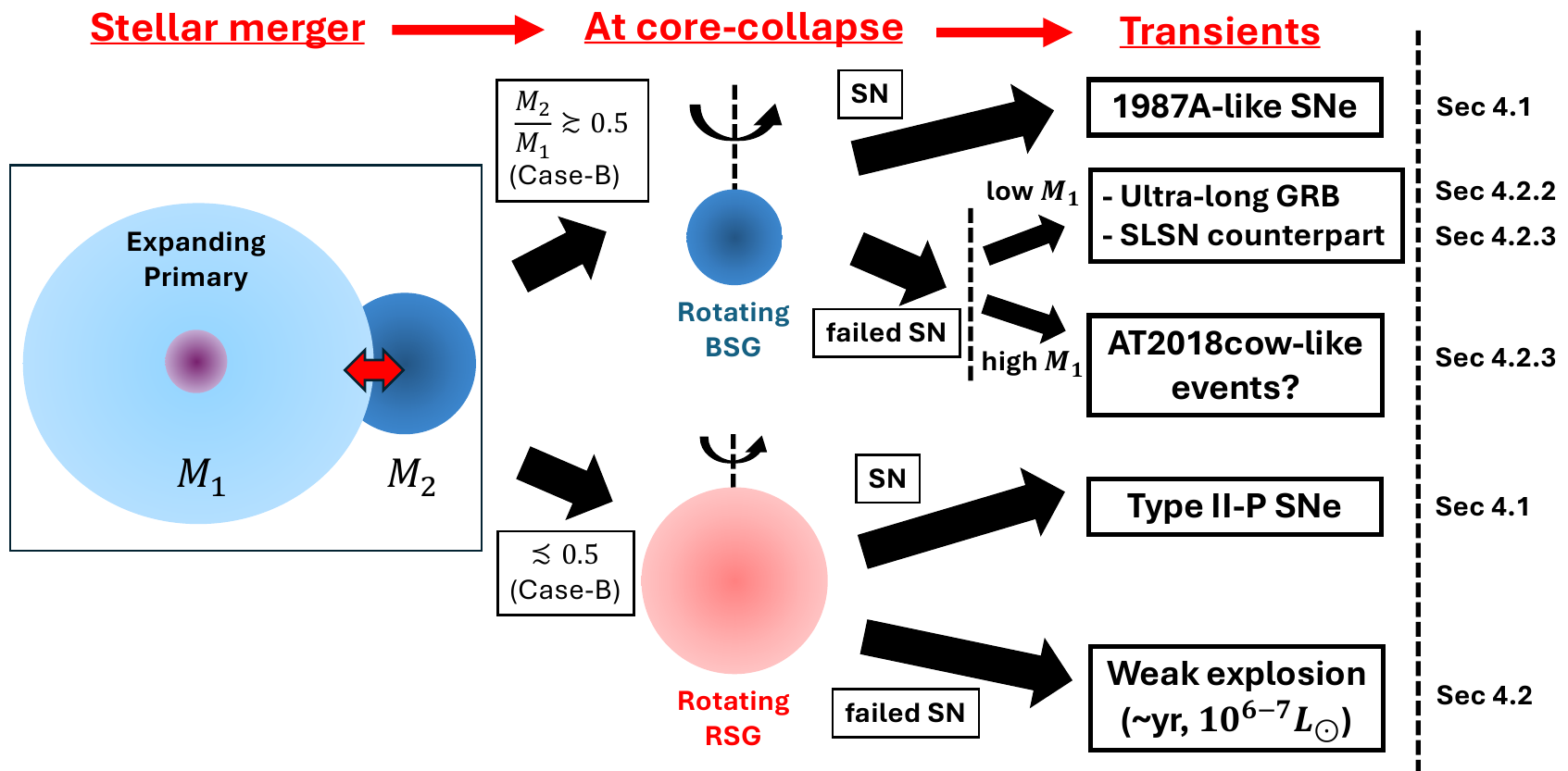}
    \caption{Summary of the transient landscape expected from our stellar models, which diverge based on the primary mass $M_1$, mass ratio of the merging stars $M_2/M_1$, and the outcome of the explosion at core-collapse (successful or failed SN). We also show the section where each of the transients is discussed.
    }
    \label{fig:schematic}
\end{figure*}

The failed SN case is found to have a richer phenomenology, in particular for BSG progenitors due to it being capable of forming an accretion disk around the newborn BH. Using a model for time-dependent accretion of the rotating envelope onto the BH, we find extreme accretion rates of $\sim 10^{-6}$--$10^{-3}M_\odot$ s$^{-1}$ over hours, with higher rates achieved for BSGs with lower ZAMS mass that maintain faster rotation. We considered the observational implications of the relativistic jet and the mildly relativistic disk wind, expected for such extreme accretion rates in the context of the collapsar scenario in GRBs. Our modeling points to these merger-origin BSGs as viable progenitors of ultra-long GRBs, as well as fast luminous transients like AT2018cow. A schematic flowchart for the transient landscape derived in this study is depicted in Figure \ref{fig:schematic}.

We conclude by raising suggestions for future improvements and applications to our modeling. First, our prescription for merger of rapid accretion onto the primary is admittedly a simplification of the various physics involved in the merger process. This most importantly neglects the details of the secondary's structure. When the mass ratio of the binary is closer to unity, the secondary is evolved in the main sequence. The secondary can then have a helium-rich core with low entropy that can preferentially sink towards the core, affecting both the core and the envelope helium abundance of the merger product. An enhanced helium abundance in the envelope leads to hotter envelopes, potentially increasing the parameter space that lead to BSGs. Recent work comparing various merger prescriptions of a single binary merger model \citep{Patton25} find differences in the core structures of the ``merger" products, while their surface properties after thermal relaxation are qualitatively similar. However the outcome should depend on the parameters governing the merger such as the mass ratio and the binary separation, and a future work of constructing a model grid with these parameters would be important.

In this work we refrained from detailed assumptions of which of the progenitors end their lives as NSs and BHs, which remains an open topic of research. Recent simulations \citep[e.g., ][Figure 15]{Burrows23} predict an ``island of BH formation" of progenitors with He core masses of $3$--$4~M_\odot$ (corresponding to ZAMS masses of $12$--$15~M_\odot$) in their stellar models, and possibly at higher ZAMS masses around $20~M_\odot$ \citep[e.g.,][]{Boccioli24}. For our model that includes convective overshoot the corresponding ZAMS mass may be slightly lower, while additional variations on explodability is expected in the merger models that experience accretion \citep[e.g.,][]{Justham14,Schneider24}. It would be interesting in future work to run our models further up to core-collapse to obtain detailed predictions of the innermost core, which could give insights on the explodability of these stars through e.g. their profiles at the silicon-oxygen interface \citep[e.g.,][]{Wang22,Boccioli23}. This requires accurately modeling (and resolving) the innermost core structures, by running with a more sophisticated nuclear reaction network with a much larger number of isotopes than that adopted in this study \citep{Farmer16,Renzo24}\footnote{The choice of nuclear networks also affected disk formation for the collapsar models considered in \cite{Renzo24}. For our BSG models at central carbon depletion, the innermost regions do not have enough AM for disk formation, and the outer edge of the carbon-oxygen core has an AM transport timescale of $r^2/\nu_{\rm AM}\sim 10^4$ yrs, much longer than the time left till core-collapse. We thus expect that evolving the BSG models further would not alter our predictions for the mass accretion outside the newborn BH.}.

For the BH spins obtained by the one-zone accretion disk model, we generally see an anti-correlation with the final BH mass (Figure \ref{fig:Mbh_vs_abh}). While the merger products from field binaries would lead to an isolated single BH, those in dynamical environments can further interact and merge with other compact objects. Massive star mergers can occur frequently in the centers of clusters, if they gather in the center upon formation by (primordial) mass segregation. They are potentially important for producing massive mass-gap BHs \citep[e.g.,][]{PortegiesZwart04,Kremer20,Gonzalez21} and spinning BHs observed in some gravitational-wave events, which are generally difficult to produce by multiple generations of BH mergers due to the large recoil kicks expected during the hierarchical merger process. Investigations of the final masses and spins of stellar merger products, using an expanded grid of mass and metallicity, would be an important piece to better understand such channels for very massive and/or spinning BHs.

\section*{Acknowledgements}
D.T. is supported by the Sherman Fairchild Postdoctoral Fellowship at Caltech. We thank Samantha Wu, Ryan Chornock, Kaustav Das, Kazumi Kashiyama, Kyle Kremer, Anthony Piro, Toshikazu Shigeyama, Mathieu Renzo, and David Vartanyan for valuable discussions. Finally, we deeply thank Bill Paxton for his continuous developments of the MESA code, without which this work and many other of our works have not been possible -- the MESA code will live on and continue to shine.



\bibliography{reference} 
\bibliographystyle{aasjournal}



\appendix 

\section{Jet propagation in BSG Envelope}
\label{app:jet_propagation}
A semi-analytical model for the interaction of a relativistic jet with the surrounding stellar envelope was provided by \cite{Bromberg11}. Here we adopt this model for the propagation of the accretion-powered jet through the BSG envelope, to estimate when/if the jet can break out from the star. The jet power is set as
\begin{eqnarray}
    L_{\rm jet}(t) = \frac{1}{2}\eta_{\rm jet}\dot{M}_{\rm BH}(t)c^2
\end{eqnarray}
where $\eta_{\rm jet}=0.01$ (see Section \ref{sec:ulGRBs}) is the jet efficiency, and here a factor $1/2$ is included as we consider only one side of the jet (towards the observer). A top-hat jet with opening angle of $\theta_0$ at launch, drills through the stellar material with a density profile $\rho_*(r)$. We set $\theta_0=0.3$ rad following \cite{Perna18}, which was motivated by values constrained from ultra-long GRBs \citep[e.g.,][]{Levan14,Kann18}.

We assume the jet launched at $t=0$ travels unimpeded (with velocity $c$) out to the innermost radius $r_{\rm in, circ}$ where the fallback material can circularize, and starts interacting with the BSG material at time $t=r_{\rm in, circ}/c$ from launch. The initial radius of the jet head and cocoon is set as $r_{h,0}=r_{\rm in, circ}$ and $r_{c,0}=(\tan\theta_0)r_{\rm in, circ}$ respectively. For simplicity, we assume the BSG envelope as static and neglect the potential infall of the BSG envelope exterior to $r_{\rm in, circ}$.

At a given time, the velocity of the jet's head is set by pressure balance as \citep{Matzner03}
\begin{eqnarray}
    \frac{dr_{\rm h}}{dt} = \frac{N_s c}{1+\tilde{L}^{-1/2}}
\end{eqnarray}
where $\tilde{L}$ is a dimensionless parameter
\begin{eqnarray}
    \tilde{L}(t) = \frac{L_{\rm jet}(t'=t-r_{\rm h}/c)}{\Sigma_{\rm jet}\rho_*(r=r_{\rm h}) c^3},
\end{eqnarray}
$N_s$ is an order-unity factor derived from calibration to simulations \citep[eq 13 of][]{Harrison18}, and $\Sigma_{\rm jet}$ is the jet's cross section set by whether the jet is collimated by the cocoon's pressure $P_c$ as derived below. The jet continuously dissipates its energy into the cocoon, and neglecting adiabatic losses \citep[see Appendix of][]{Bromberg11} the cocoon's energy (focusing only on the observer's side) is
\begin{eqnarray}
    E_c(t) \approx \int_0^{t-r_h/c} dt' L_{\rm jet}(t')
\end{eqnarray}
Due to its pressure the cocoon expands sideways with velocity
\begin{eqnarray}
    \frac{dr_{\rm c}}{dt} = \sqrt{\frac{P_c}{\bar{\rho}_*}},
\end{eqnarray}
where $\bar{\rho}_*$ is the mean density in the volume occupied by the cocoon
\begin{eqnarray}
    \bar{\rho}_* = \frac{M_c}{V_c} \approx  \frac{\int_{r_{h,0}/c}^{t} dt' [\pi r_c^2 \rho_*v_h]}{\int_{r_{h,0}/c}^{t} dt' [\pi r_c^2v_h]},
\end{eqnarray}
where $V_c$ and $M_c$ are respectively the volume and mass of the cocoon.
The pressure of the cocoon is dominated by radiation
\begin{eqnarray}
    P_c(t) = \frac{E_c}{3V_c}.
\end{eqnarray}
The collimation shock by the cocoon splits the jet into an unshocked region and shocked region. The unshocked jet's height is given as a function of $L_{\rm jet}$ and $P_c$ as \citep[][their Figure 2]{Bromberg11} 
\begin{eqnarray}
    \hat{z} \approx \sqrt{\frac{L_{\rm jet}}{\pi c P_c}}
\end{eqnarray}
with the jet being uncollimated below $\approx \hat{z}/2$ and collimated above. Comparison of $\hat{z}/2$ and $r_h$ gives the criterion of collimation at the jet head, and the jet cross section at the head is then
\begin{eqnarray}
    \Sigma_{\rm jet} \approx  
    \begin{cases}
       \pi \theta_0^2 r_h^2 & ({\rm uncollimated}) \\
       \pi \theta_0^2 (\hat{z}/2)^2 = L_{\rm jet}\theta_0^2/(4cP_c) & ({\rm collimated}).
    \end{cases}
\end{eqnarray}

The jet breaks out the star when the jet head $r_h$ reaches the stellar radius $R_*$, and we define the time this occurs as the breakout time $t_{\rm bo}$ in Section \ref{sec:ulGRBs}. 
We note that the model is not strictly applicable close to the stellar surface, where the density profile $\rho_*(r)$ sharply drops (faster than $r^{-3}$). The forward shock accelerates and loses causal contact with the jet, and the assumption of pressure balance can break down \citep{Bromberg11}. Comparison of the analytical model with hydrodynamical simulations, for a GRB-like jet propagating through a Wolf-Rayet star, nevertheless shows reasonable agreement for the estimation of $t_{\rm bo}$ \citep{Harrison18}.

\section{Model variations}
\label{app:model_variations}

\begin{figure*}
   \centering  \includegraphics[width=\linewidth]{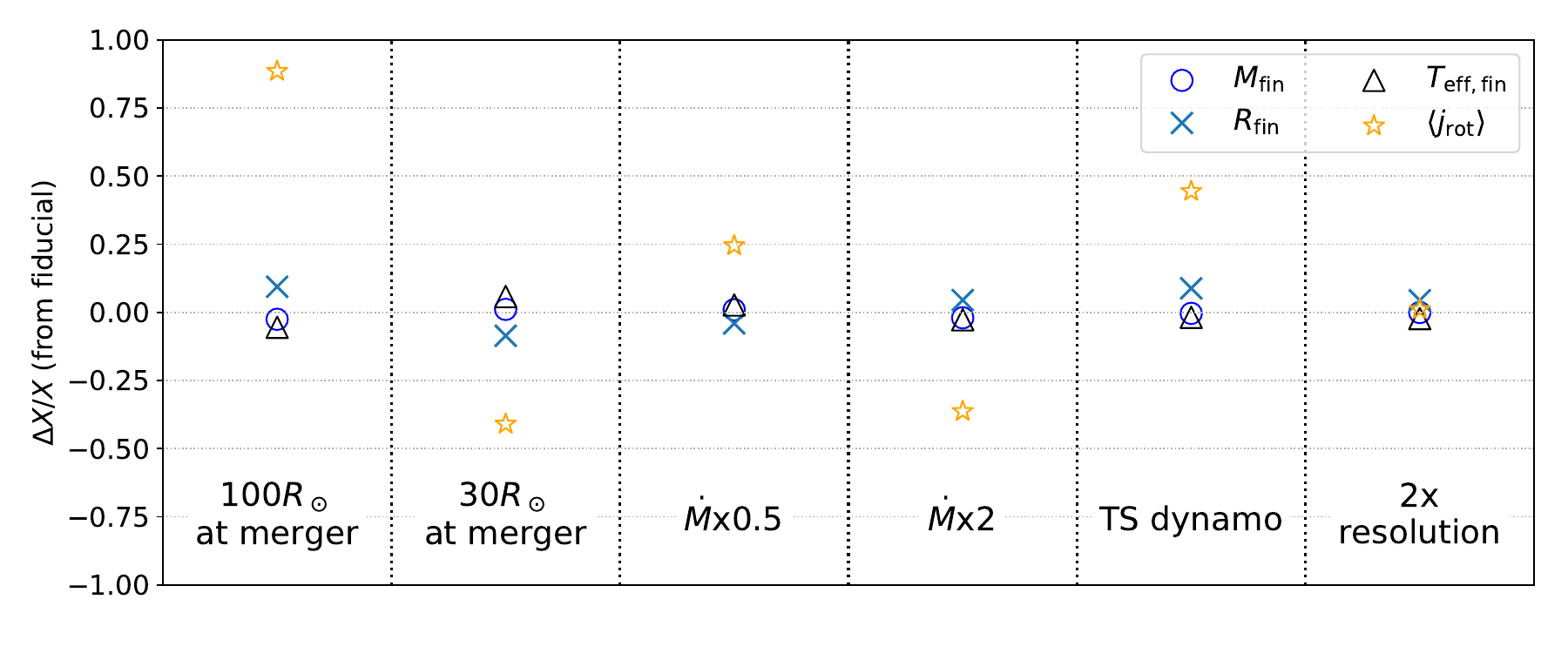}
    \caption{Fractional difference in the stellar properties most relevant to our paper, for variations in the input physics. We have chosen the fiducial BSG model of $M_1=15~M_\odot$, $M_2=8~M_\odot$ that is used in other plots.}
    \label{fig:variations}
\end{figure*}

To check how the results are sensitive to various prescriptions, we re-run our baseline BSG model of $M_1=15~M_\odot$, $M_2=8~M_\odot$ with variations in the input physics as follows: 
\begin{enumerate}
    \item we change the onset of mass accretion from a primary radius from our fiducial value of $50~R_\odot$ to [$30~R_\odot, 100~R_\odot$].
    \item we change the wind efficiency (\texttt{Dutch\_scaling\_factor}) that controls the mass loss rate, by a factor 2 on both sides from our fiducial value of $0.5$.
    \item we update the AM transport scheme from the \cite{Fuller19} formalism to the default prescription based on the Taylor-Spruit (TS) dynamo \citep{Spruit02}.
    \item we increase the mass resolution by a factor of 2, by lowering two parameters in MESA \texttt{mesh\_delta\_coeff} and \texttt{mesh\_delta\_coeff\_for\_highT} from our default 0.8 to 0.4.
\end{enumerate}
Figure \ref{fig:variations} shows the differences in four parameters of the pre-SN star most relevant for our work -- mass, stellar radius, effective temperature, and mass-averaged specific AM  ($M_{\rm fin}, R_{\rm fin}, T_{\rm eff, fin}$ and $\langle j_{\rm rot}\rangle$ in Table \ref{tab:summary}). We find the stellar parameters ($M_{\rm fin}, R_{\rm fin}, T_{\rm eff, fin}$) to weakly vary within $10\%$, suggesting our results and qualitative picture of the RSG/BSG dichotomy are not subject to these uncertainties. The largest variations are seen in the mass-averaged AM $\langle j_{\rm rot}\rangle$, with larger radii at merger increasing the AM and larger mass loss rate decreasing the AM, both expected from our framework of AM evolution (equations \ref{eq:j_acc}, \ref{eq:dJdt}). The effects of AM prescription are less clear, but the larger AM content for the TS dynamo case may be explained by a smaller AM loss by winds, due to less efficient AM transport to the surface. While they are important model uncertainties, we expect them to not change our qualitative picture of the transient landscape expected from these stars.

Finally, our parameters are unchanged within $5\%$ when the mass resolution is increased by a factor of 2, and we consider the models to be sufficiently converged in resolution for our purpose.

\section{Summary of Model Grid}
In Table \ref{tab:summary} we show our model grid and the resulting stellar properties at core-collapse.
\begin{table*}
    \centering
    \caption{Summary of our 32 models of single and merger-remnant stars computed by MESA and their post-main sequence properties.}
    \begin{tabular}{cc|ccccccc}
    $M_{1,\rm ZAMS}$ [$M_\odot$] & $M_2$ [$M_\odot$] & $M_{\rm fin}$ [$M_\odot$] & $M_{\rm He, fin}$ [$M_\odot$] & $R_{\rm fin}$ [$R_\odot$] & $T_{\rm eff,Y_{\rm He}=0.5}$ [K] & $T_{\rm eff, fin}$ [K] & $\langle j_{\rm rot}\rangle$ [$10^{17}\ {\rm cm^2\ s^{-1}}$] & $M_{j>j_{\rm ISCO, Sch}}$ [$M_\odot$]\\
    \hline
10 & 0 & 9.6 & 3.0 & 658 & 3720 & 3337 & 1.90 & 5.28\\
10 & 2 & 11.1 & 3.0 & 644 & 3768 & 3387 & 2.97 & 6.92\\
10 & 4 & 12.9 & 3.3 & 678 & 3808 & 3410 & 3.64 & 8.33\\
10 & 6 & 14.9 & 3.5 & 698 & 19116 & 3457 & 3.55 & 9.56\\
{\bf 10} & {\bf 8} & {\bf 16.6} & {\bf 3.6} & {\bf 34} & {\bf 21207} & {\bf 15466} & {\bf 3.51} & {\bf 8.42} \\ \hline
15 & 0 & 13.6 & 4.8 & 992 & 3572 & 3318 & 1.92 & 6.38\\
15 & 2 & 15.0 & 4.9 & 997 & 3578 & 3339 & 1.64 & 6.27\\
15 & 4 & 17.8 & 5.4 & 1050 & 14570 & 3363 & 1.64 & 6.70\\
15 & 6 & 19.6 & 5.2 & 971 & 19776 & 3436 & 2.11 & 8.52\\
{\bf 15} & {\bf 8} & {\bf 21.4} & {\bf 5.1} & {\bf 50} & {\bf 21805} & {\bf 15128} & {\bf 2.13} & {\bf 5.47} \\
{\bf 15} & {\bf 10} & {\bf 23.2} & {\bf 4.9} & {\bf 36} & {\bf 23568} & {\bf 18761} & {\bf 2.03} & {\bf 5.41} \\
{\bf 15} & {\bf 12} & {\bf 25.2} & {\bf 4.7} & {\bf 29} & {\bf 25156} & {\bf 21367} & {\bf 2.09} & {\bf 5.69} \\ \hline
20 & 0 & 17.7 & 7.2 & 1346 & 8775 & 3318 & 0.05 & 0.00\\
20 & 2 & 19.9 & 7.3 & 1338 & 12047 & 3341 & 0.11 & 0.00\\
20 & 4 & 22.2 & 7.4 & 1324 & 15154 & 3370 & 0.36 & 0.00\\
20 & 6 & 24.4 & 7.2 & 1254 & 18800 & 3424 & 0.99 & 1.05\\
{\bf 20} & {\bf 8} & {\bf 26.1} & {\bf 6.8} & {\bf 139} & {\bf 20811} & {\bf 10356} & {\bf 1.01} & {\bf 1.80} \\
{\bf 20} & {\bf 10} & {\bf 27.9} & {\bf 6.7} & {\bf 81} & {\bf 22445} & {\bf 13960} & {\bf 1.06} & {\bf 1.93} \\
{\bf 20} & {\bf 12} & {\bf 29.7} & {\bf 6.4} & {\bf 59} & {\bf 23966} & {\bf 16809} & {\bf 1.11} & {\bf 1.89} \\
{\bf 20} & {\bf 14} & {\bf 31.7} & {\bf 6.5} & {\bf 51} & {\bf 25283} & {\bf 18434} & {\bf 1.23} & {\bf 2.07} \\
{\bf 20} & {\bf 16} & {\bf 33.7} & {\bf 6.2} & {\bf 45} & {\bf 26458} & {\bf 20124} & {\bf 1.28} & {\bf 1.96} \\ \hline
25 & 0 & 20.6 & 9.5 & 1602 & 8354 & 3338 & 0.06 & 0.00\\
25 & 2 & 23.4 & 9.5 & 1593 & 11224 & 3355 & 0.07 & 0.00\\
25 & 4 & 26.4 & 9.5 & 1563 & 14188 & 3385 & 0.08 & 0.00\\
25 & 6 & 28.8 & 9.3 & 1502 & 16911 & 3427 & 0.19 & 0.00\\
25 & 8 & 30.6 & 9.0 & 1460 & 19031 & 3454 & 0.45 & 0.00\\
25 & 10 & 32.4 & 8.8 & 445 & 20681 & 6436 & 0.41 & 0.35\\
{\bf 25} & {\bf 12} & {\bf 34.1} & {\bf 8.6} & {\bf 219} & {\bf 21995} & {\bf 9400} & {\bf 0.48} & {\bf 0.23} \\
{\bf 25} & {\bf 14} & {\bf 35.9} & {\bf 8.5} & {\bf 158} & {\bf 23326} & {\bf 11256} & {\bf 0.53} & {\bf 0.22} \\
{\bf 25} & {\bf 16} & {\bf 38.0} & {\bf 8.2} & {\bf 131} & {\bf 24380} & {\bf 12559} & {\bf 0.65} & {\bf 0.33} \\
{\bf 25} & {\bf 18} & {\bf 40.0} & {\bf 8.3} & {\bf 108} & {\bf 25389} & {\bf 14103} & {\bf 0.73} & {\bf 0.40} \\
{\bf 25} & {\bf 20} & {\bf 42.0} & {\bf 8.3} & {\bf 89} & {\bf 26275} & {\bf 15898} & {\bf 0.80} & {\bf 0.42}
    \end{tabular}
    \vspace{2mm}\\
    {Note: Columns are the primary's ZAMS mass, the mass accreted onto the primary, final stellar mass at core C depletion, final He-core mass, final radius, effective temperatures in the middle of core He burning and at core C depletion, mass averaged specific AM, and mass of the outer layer of the final star having a specific AM higher than $j_{\rm ISCO,Sch}$ (equation \ref{eq:j_ISCO} for a Schwarzschild BH of $a_{\rm BH}=0$) The models in bold fonts are progenitors that are BSGs at core C-depletion (14 out of 32), here defined as stars with  temperatures $T_{\rm eff, fin}>10^{3.9}$ K.}
    \label{tab:summary}
\end{table*}

\end{document}